\documentclass[%
  reprint,
  amsmath,amssymb,
  aps,
  prmaterials,
  superscriptaddress
]{revtex4-2}

\usepackage{graphicx}        
\usepackage{dcolumn}         
\usepackage{bm}              
\usepackage{xspace}          
\usepackage[version=4]{mhchem} 
\usepackage{multirow}        
\usepackage{amsfonts}        
\usepackage{amsthm}          
\usepackage{mathrsfs}        
\usepackage{xcolor}          
\usepackage{textcomp}        
\usepackage{booktabs}        
\usepackage{algorithm}       
\usepackage{algorithmicx}    
\usepackage{algpseudocode}   
\usepackage{listings}        
\usepackage[caption=false]{subfig} 
\usepackage[normalem]{ulem}  
\usepackage[hidelinks]{hyperref}     
\usepackage[capitalize]{cleveref} 

\newcommand{\MBT}{MnBi\textsubscript{2}Te\textsubscript{4}\xspace}
\newcommand{\mbt}{\ce{MnBi2Te4}}
\newcommand{\mst}{\ce{MnSb2Te4}}
\newcommand{\bt}{\ce{Bi2Te3}}
\newcommand{\mt}{\ce{MnTe}}
\newcommand{\bst}{(Bi$_x$,Sb$_{1-x}$)$_2$Te$_3$}
\newcommand{\st}{\ce{Sb2Te3}}
\newcommand{\gst}{\ce{GeSb2Te4}}

\newcommand{\mnmn}{Mn$_\text{Mn}$}
\newcommand{\bimn}{Bi$_\text{Mn}$}
\newcommand{\mnbi}{Mn$_\text{Bi}$}
\newcommand{\mnte}{Mn$_\text{Te}$}
\newcommand{\bite}{Bi$_\text{Te}$}
\newcommand{\tebi}{Te$_\text{Bi}$}
\newcommand{\biteone}{Bi$_\text{Te$_1$}$}
\newcommand{\bitetwo}{Bi$_\text{Te$_2$}$}

\newcommand{\threedfive}{$3d^5$}
\newcommand{\rahe}{$R_{\text{AHE}}$}
\newcommand{\sahe}{\ensuremath{\sigma_{xy}^{A}}}
\newcommand{\sxx}{\ensuremath{\sigma_{xx}}}
\newcommand{\fluence}[2]{$#1 \times 10^{#2}$~cm$^{-2}$}
\newcommand{\fivesix}{\ce{(V)_2(VI)_3}}

\newcommand{\add}[1]{#1}
\newcommand{\del}[1]{}

\begin{document}

\title{Disorder-driven symmetry suppression by van der Waals planar
defects in a magnetic topological insulator}

\author{Rikkie Joris}
  \email{Contact author: rikkie.joris@kuleuven.be}
  \affiliation{Quantum Solid State Physics, KU Leuven,
    Celestijnenlaan 200D, B-3001 Leuven, Belgium}

\author{Heyi Xia}
  \affiliation{Department of Materials Engineering, KU Leuven,
    Kasteelpark Arenberg 44, B-3001 Leuven, Belgium}

\author{Ana Beatriz Pedro Fontes}
  \affiliation{Institute of Condensed Matter and Nanosciences,
    Université Catholique de Louvain,
    B-1348 Louvain-la-Neuve, Belgium}

\author{Sara Bey}
  \affiliation{Department of Physics, University of Notre Dame,
    Notre Dame, Indiana 46556, USA}

\author{Seul-Ki Bac}
  \affiliation{Institut des NanoSciences de Paris, INSP,
    Sorbonne Université, CNRS,
    4 place Jussieu, F-75005 Paris, France}

\author{Jiaqi Zhou}
  \affiliation{Institute of Condensed Matter and Nanosciences,
    Université Catholique de Louvain,
    B-1348 Louvain-la-Neuve, Belgium}

\author{Zviadi Zarkua}
  \affiliation{Quantum Solid State Physics, KU Leuven,
    Celestijnenlaan 200D, B-3001 Leuven, Belgium}

\author{Ahmed Samir Lotfy}
  \affiliation{Quantum Solid State Physics, KU Leuven,
    Celestijnenlaan 200D, B-3001 Leuven, Belgium}

\author{Muhammad Saad}
  \affiliation{Quantum Solid State Physics, KU Leuven,
    Celestijnenlaan 200D, B-3001 Leuven, Belgium}

\author{Philippe Ohresser}
  \affiliation{Synchrotron SOLEIL, L'Orme des Merisiers,
    F-91190 Saint Aubin, France}

\author{Bence G. M\'arkus}
  \affiliation{Department of Physics, University of Notre Dame, Notre Dame, Indiana 46556, USA}
  \affiliation{Stavropoulos Center for Complex Quantum Matter, University of Notre Dame, Notre Dame, Indiana 46556, USA}

\author{L\'aszl\'o Forr\'o}
  \affiliation{Department of Physics, University of Notre Dame, Notre Dame, Indiana 46556, USA}
  \affiliation{Stavropoulos Center for Complex Quantum Matter, University of Notre Dame, Notre Dame, Indiana 46556, USA}

\author{Margriet van Bael}
  \affiliation{Quantum Solid State Physics, KU Leuven,
    Celestijnenlaan 200D, B-3001 Leuven, Belgium}

\author{Clement Merckling}
  \affiliation{Department of Materials Engineering, KU Leuven,
    Kasteelpark Arenberg 44, B-3001 Leuven, Belgium}
  \affiliation{IMEC, B-3001 Leuven, Belgium}

\author{Xinyu Liu}
  \affiliation{Department of Physics, University of Notre Dame,
    Notre Dame, Indiana 46556, USA}

\author{Francisco Molina-Lopez}
  \affiliation{Department of Materials Engineering, KU Leuven,
    Kasteelpark Arenberg 44, B-3001 Leuven, Belgium}

\author{Jean-Christophe Charlier}
  \affiliation{Institute of Condensed Matter and Nanosciences, Université Catholique de Louvain,
    B-1348 Louvain-la-Neuve, Belgium}

\author{Jin Won Seo}
  \affiliation{Department of Materials Engineering, KU Leuven,
    Kasteelpark Arenberg 44, B-3001 Leuven, Belgium}

\author{Badih A. Assaf}
  \affiliation{Department of Physics, University of Notre Dame, Notre Dame, Indiana 46556, USA}

\author{Lino M. C. Pereira}
  \affiliation{Quantum Solid State Physics, KU Leuven,
    Celestijnenlaan 200D, B-3001 Leuven, Belgium}

\date{\today}

\begin{abstract}
Magnetic topological insulators offer a platform to control electronic topology through magnetic order, yet reliable routes to tune their properties remain limited. Here, we show that ion irradiation allows to modify the magnetic properties and the \del{topological}\add{anomalous Hall response} of the van der Waals magnetic topological insulator MnBi$_2$Te$_4$. Using noble gas ion beams, intrinsic defects are introduced via collision cascades without chemical doping. We identify two distinct regimes. At low fluence, cation antisite disorder leads to a near-complete redistribution of Bi over cation sites while preserving long-range crystallographic order, accompanied by a transition from $p$-type to $n$-type transport. At high fluence, cation–anion intermixing drives the formation of a previously unreported layer-disordered phase characterized by a high density of van der Waals–specific planar defects, including swapped bilayers. Despite significant structural disorder, the system retains partial periodic order up to high displacement levels. Magnetometry and X-ray spectroscopy show that the Mn high-spin state and \add{intralayer} antiferromagnetic interactions\add{\ likely} persist, while \add{macroscopic} magnetic anisotropy is strongly reduced. At the same time, the \add{magnitude of the} anomalous Hall conductivity is \del{suppressed fivefold} \add{reduced by an order of magnitude}, far exceeding the change in magnetization, \add{and its sign is modified.} \add{The observations are consistent with a suppression of the Berry curvature of the occupied states.}\del{indicating a direct modification of Berry curvature.} These results establish ion irradiation as a means to tune \del{topology}\add{anomalous Hall conductivity in topological insulators} through defect engineering and reveal a disorder-driven approach to control symmetry and electronic structure in van der Waals magnetic materials.
\end{abstract}

\maketitle

\section*{Introduction}\label{introduction}

Magnetic topological insulators (MTIs) have emerged as a fertile area of research due to the interplay between magnetism and band topology that results in emerging quantum phenomena~\cite{hasan2010,ando2013,qi2011,yu2010,chang2013}. Examples include the quantum anomalous Hall effect (QAHE)~\cite{yu2010,chang2013,chang2015}, Weyl semimetals~\cite{soluyanov2015}, axion states~\cite{mogi2017}, and Majorana fermions~\cite{qi2011,sato2017,uday2024}, with applications ranging from dissipationless edge currents to quantum computation~\cite{tokura2019}. The search for materials hosting both band inversion and magnetic ordering has been successful in the transition metal (TM) doped tetradymite-type materials \ce{(V)_2(VI)_3} with the group V elements Bi/Sb and the group VI elements Te/Se~\cite{yu2010,chang2013,chang2015,ren2010,zhang2009,heremans2017}. The disordered ferromagnet V/Cr-doped \bst\ (V/CBST) and ordered antiferromagnet \mbt\ (MBT)~\cite{zhang2019,li2019,otrokov2019,gong2019,he2020} have proven to be fruitful material platforms for the exploration of topological phases of matter under time reversal symmetry breaking. Fine control over the magnetic order and electronic properties in both systems remains challenging, in particular the tuning of the Fermi level into the exchange gap of the topological surface state. The inherent charge puddling in V/CBST~\cite{lippertz2022} and the lack of a reliable approach to control the Fermi level in MBT~\cite{yan2022, wang2025,bai2024} continue to hinder progress toward Majorana physics in these systems. Although the QAHE has been observed in odd-layer MBT films due to uncompensated magnetization~\cite{deng2020}, the antiferromagnetic nature of the material limits its potential as a platform for the exploration of Weyl physics for which a ferromagnetic phase is required~\cite{li2019}. Control over cation antisite defects is a key objective in both systems, and in MBT specifically, given their influence on the magnetic order and transport properties~\cite{yan2022}. 
Ion irradiation has been widely used to tune the Fermi level in semiconductors and to control the magnetic properties of metal thin films~\cite{lugakov1982,yuan2022,fassbender2004}. When performed with inert ions, this process predominantly generates intrinsic point defects through collision cascades, rather than introducing dopants, thereby enabling controlled modification of material properties without altering chemical composition. However, the mechanisms governing ion irradiation in tetradymite and van der Waals (vdW) materials remain poorly understood~\cite{long2023,abhirami2021,fominykh2026}. While ion irradiation and implantation have proven effective in tuning the transport and thermoelectric properties of the \ce{(V)_2(VI)_3} class, a microscopic understanding of the chemical and structural changes induced by ion bombardment in vdW materials remains largely lacking, beyond phenomenological descriptions of damage accumulation and amorphisation~\cite{abhirami2024,cortie2020}. In this work, we identify a regime where self-assembly of cation-anion antisites generated by ion bombardment results in the formation of a novel disordered phase where planar symmetry in the vdW material is partially suppressed. The discovery of this pathway to resist amorphisation, whereby ion beam damage is dispersed across multiple antisites that self-organize into higher-dimensional defect structures, introduces a new category of radiation tolerance in crystalline materials. \del{The key factor is that the generated planar defects produce a partially disordered crystal where some symmetries are inherited and others suppressed.} \add{The key factor is that the generated planar defects produce a partially disordered crystal in which the periodicity along the $c$-axis is suppressed, while periodic order along other crystallographic directions is retained.} A detailed understanding of these mechanisms is essential, as beam–solid interactions may provide a route to systematically control the magnetic and transport properties of existing MTIs\add{, as has been established for other material classes such as the oxide-based quantum materials \cite{herklotz2025}}. 

Beyond conventional electronic and magnetic parameters, the topological nature of MTIs introduces the Berry curvature as an additional degree of freedom. In this work, we investigate the effects of ion irradiation on \mbt\ thin films and show that \del{Berry curvature can be tuned (suppressed)} \add{the anomalous Hall conductivity can be tuned} through ion bombardment, without inducing amorphisation of the structure.
Two regimes are observed, a low fluence regime where cation antisite disorder changes the magnetic and transport properties of the material while leaving the crystalline structure essentially unchanged, and a high fluence regime where cation-anion intermixing results in a novel disordered phase that contains a high density of two-dimensional (2D) defects. This disordered crystal structure arises due to the stability of planar defects specific to vdW materials, most notably swapped bilayers and ripplocations\add{, defects where a bilayer detaches from a vdW layer and joins another or the same vdW layer, respectively}. Such defects have been observed in related compounds, namely \ce{(V)_2Te3} with V = Bi/Sb and \gst, where they are formed due to cation-anion site disorder introduced during growth or after mechanical deformation~\cite{li2025,wang2019,wang2021,deng2024,wang2018}. Already in the lower-fluence (cation disorder) regime, we observe a \del{fivefold} \add{fourfold} suppression of the anomalous Hall conductivity without a dramatic reduction of the magnetic moment, which \del{we tie} \add{can be linked} to the tuning of Berry curvature. In the disordered crystal that emerges in the high fluence regime, the degradation of the anomalous Hall conductivity persists while longitudinal conductivity \del{drastically} increases.

To elucidate the structural origin of these effects, we perform a detailed experimental investigation of the resulting 2D defects using scanning transmission electron microscopy (STEM) and energy-dispersive X-ray spectroscopy (EDS). We show that the formation of these planar defects can be understood at the atomic level as \bite\ and \tebi\ antisite defects that agglomerate into complex microstructures, as evidenced by density functional theory (DFT) calculations and large-scale relaxations performed using a MACE machine learning potential within the Atomic Simulation Environment.
\del{With the comprehension of the mechanics responsible for the disordered phase, we discuss other vdW materials that may host such effects, and to this end show that X-ray Diffraction (XRD) can be used to identify the correct fluence regimes from macroscopic measurements.} Finally, we analyze the effects on the magnetic and transport properties of \mbt\ (with a focus on the low fluence regime) and discuss the viability of ion irradiation as a method of tuning the properties of this family of MTIs. 

\section*{Results}\label{results}

An epitaxial MBT film (60 nm) was grown on a \add{2-inch} \ce{GaAs}(111) wafer with a \bt\ (BT) buffer \add{(to improve epitaxy)} using molecular beam epitaxy (MBE). The film was irradiated with a 300 keV Ar$^{2+}$ beam, such that the incorporation of Ar as an impurity is negligible, enabling the study of disorder arising solely from collision cascades initiated by the incident ions. The wafer was diced into separate pieces, which were then irradiated at various fluences. In this work, we focus on \fluence{5}{13}\ and \fluence{2}{14}, referred to as low and high fluence, respectively, throughout the text. The analysis of the STEM and EDS data focuses on the central region of the MBT film, where surface and interface effects can be neglected, allowing us to gain clearer insight into the intrinsic effect of the collision cascades in MBT. According to our SRIM-2013 simulations~\cite{ziegler2010srim}, each atom in this region of the film is displaced $0.17$ and $0.66$ times on average for the low and high fluence, respectively. \add{SRIM assumes a static and homogeneous target with binary collisions by projectile atoms. The simulated displacement and ion ranges can therefore differ from the true values, but still represent a realistic order of magnitude estimate.} Further details on the irradiation parameters (such as the displacements per atom) and an overview of the underlying physics are provided in the Supplemental Material~\cite{SM}.

\begin{figure*}
    \includegraphics[width=\textwidth]{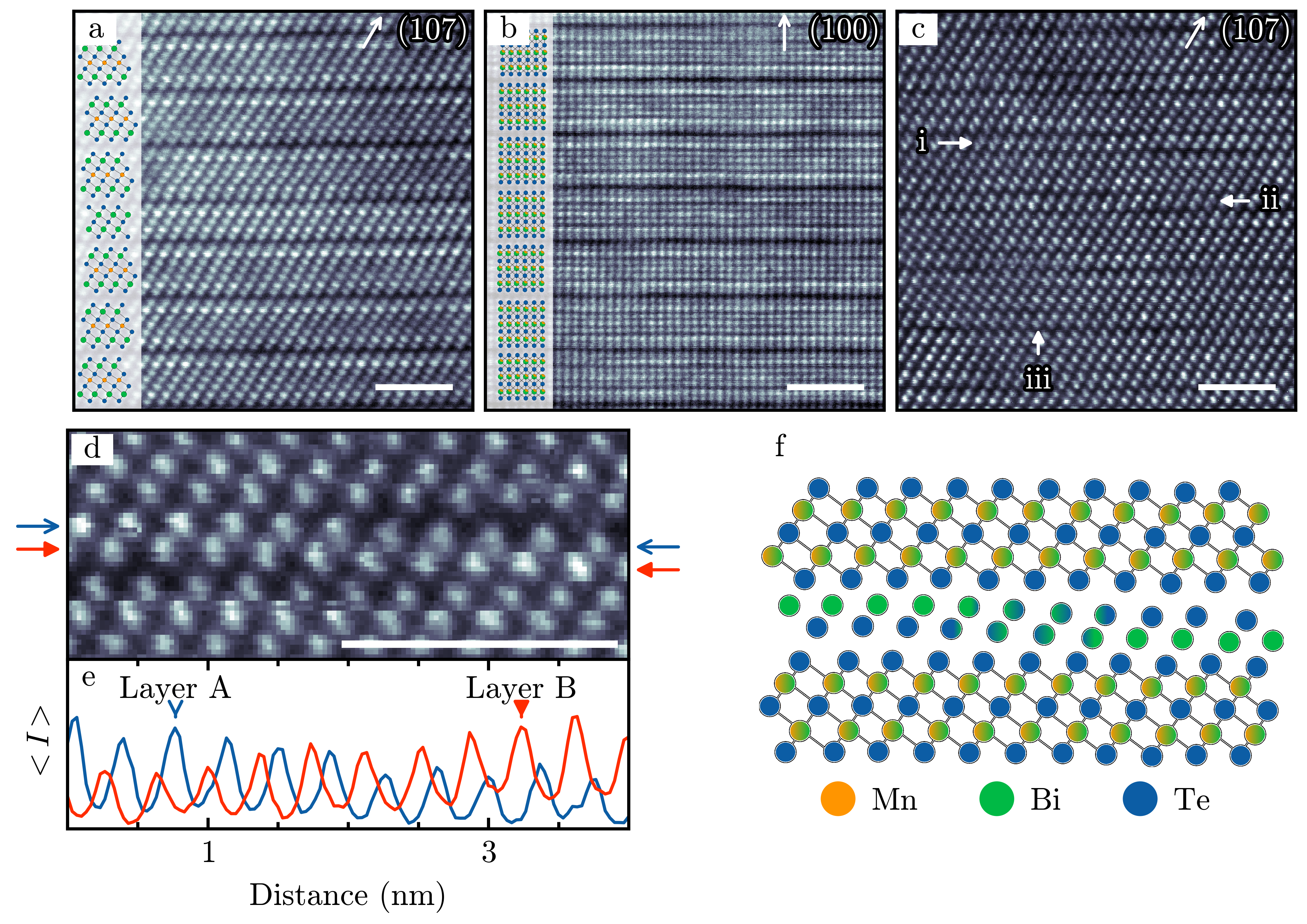}
    \caption{HAADF-STEM micrographs of (\textbf{a}) pristine, (\textbf{b}) lightly irradiated and (\textbf{c}) highly irradiated \mbt\ films. \add{Schematics of the observed structure are given on the left in (\textbf{a}, \textbf{b}).} Annotated on the high fluence irradiated micrograph are (i) a swapped bilayer, (ii) a detached bilayer stable in the vdW gap, and (iii) a ripplocation. Relevant crystallographic directions are given in the top right. (\textbf{d}) HAADF-STEM image of a swapped bilayer with (\textbf{e}) the average intensity of the layers within the bilayer over the lateral distance. (\textbf{f}) A diagrammatic representation of a swapped bilayer defect. The site occupancy and bonds indicated are schematic representations intended to guide the eye and should not be interpreted as quantitatively accurate. The chemical nature of the quintuple and bilayers is discussed in the main text. All scale bars are 2 nm.}
    \label{fig:tem-img}
\end{figure*}

Before irradiation (in the pristine state), the crystal structure of the MBT film consists of unit cells containing three septuple layers (SLs) stacked in ABC fashion, each SL containing (Te$_1$-Bi-Te$_2$-Mn-Te$_2$-Bi-Te$_1$) planes~\cite{zhang2019,li2019,otrokov2019,gong2019}. Figure \ref{fig:tem-img}(a) shows a representative atomic resolution high-angle annular dark field (HAADF)-STEM micrograph of the as-prepared film, in which the SL ordering and a well-defined vdW gap are clearly visible. The Bi basal planes can be readily identified by their brighter atomic columns due to the larger atomic number of Bi, as the contrast in HAADF imaging scales approximately with the atomic number (Z-contrast). By comparison, the Mn and Te planes exhibit lower contrast, with the Mn atomic columns displaying the weakest intensity. In the middle of the micrograph, an intercalated BT layer can be observed (a common growth defect~\cite{lapano2020,kagerer2020,inoch2025,he2023,luo2023}) which appears with a $\sim 1:7$ (BT : MBT) concentration in the pristine film. After low-fluence irradiation, the SL structure remains intact and retains well defined vdW gaps (Fig. \ref{fig:tem-img}(b)). The loss of \add{(local)} contrast between Bi and Mn planes indicates the presence of chemical disorder as discussed below.

High-fluence irradiation results in a novel defective structure where the long-range crystallographic order along the $c$-axis of the vdW material is heavily affected (Fig. \ref{fig:tem-img}(c)). The terminating bilayers of the SLs detach from the host layer, leading to one of the following outcomes: they rejoin the original layer (ripplocation~\cite{kushima2015,deng2024}); attach to an adjacent vdW layer (swapped bilayer~\cite{deng2024}); or remain stable within the vdW gap over extended lateral distances. \add{We assume the latter is either a ripplocation or swapped bilayer that joins an adjacent vdW layer outside of the domain of the micrograph.} These planar defects manifest as dislocations along the $a$- or $b$-axis, while the $(107)$ direction remains largely free of dislocations, resulting in a partially preserved periodic order. This selective suppression of crystallographic symmetry is further discussed in the Supplemental Material~\cite{SM}. We note an absence of vdW layers containing 3 or 9 atomic layers in this disordered structure. The MBT SL either remains locally intact or splits into a quintuple layer (QL) and a bilayer (BL). This implies that swapped bilayers and ripplocations are not independent but correlated over multiple unit cells, since random detachment and swapping of BLs would lead to the formation of 3- and 9-layered structures.

To understand the nature and formation mechanism of the detaching bilayers, their elemental composition must be identified. Firstly, we establish the role of Bi in a swapped bilayer defect by tracking Z-contrast-enhanced atomic columns across the bilayer. Secondly, the elemental composition of a ripplocation defect is extracted from spatially resolved EDS curves. Figures \ref{fig:tem-img}(d,f) contain a diagram and STEM micrograph of a swapped bilayer defect. The bilayer swaps between two quintuple layers over a range of several nm to locally form SLs on the left and right hand side of the image. The average intensity of the top (A) and bottom (B) layers of the swapped bilayer are plotted as functions of lateral distance in Figure \ref{fig:tem-img}(e). Layer B contains atomic columns with lower intensity than layer A on the left hand side (LHS) of the micrograph but these increase in intensity during and after the crossing of the vdW gap. Layer A has the opposite behavior. This corresponds to the interpretation that layer A starts (LHS) as the Bi (cation) layer of the top SL and evolves to be the terminating Te (anion) layer in the bottom SL. Identical observations have been made in the sister compound \gst\ and the parent compounds \bt\ and \st\ where swapped bilayers are known to form due to cation-anion antisites~\cite{wang2019,wang2021,deng2024,wang2018}.

\begin{figure*}
    \includegraphics[width=\textwidth]{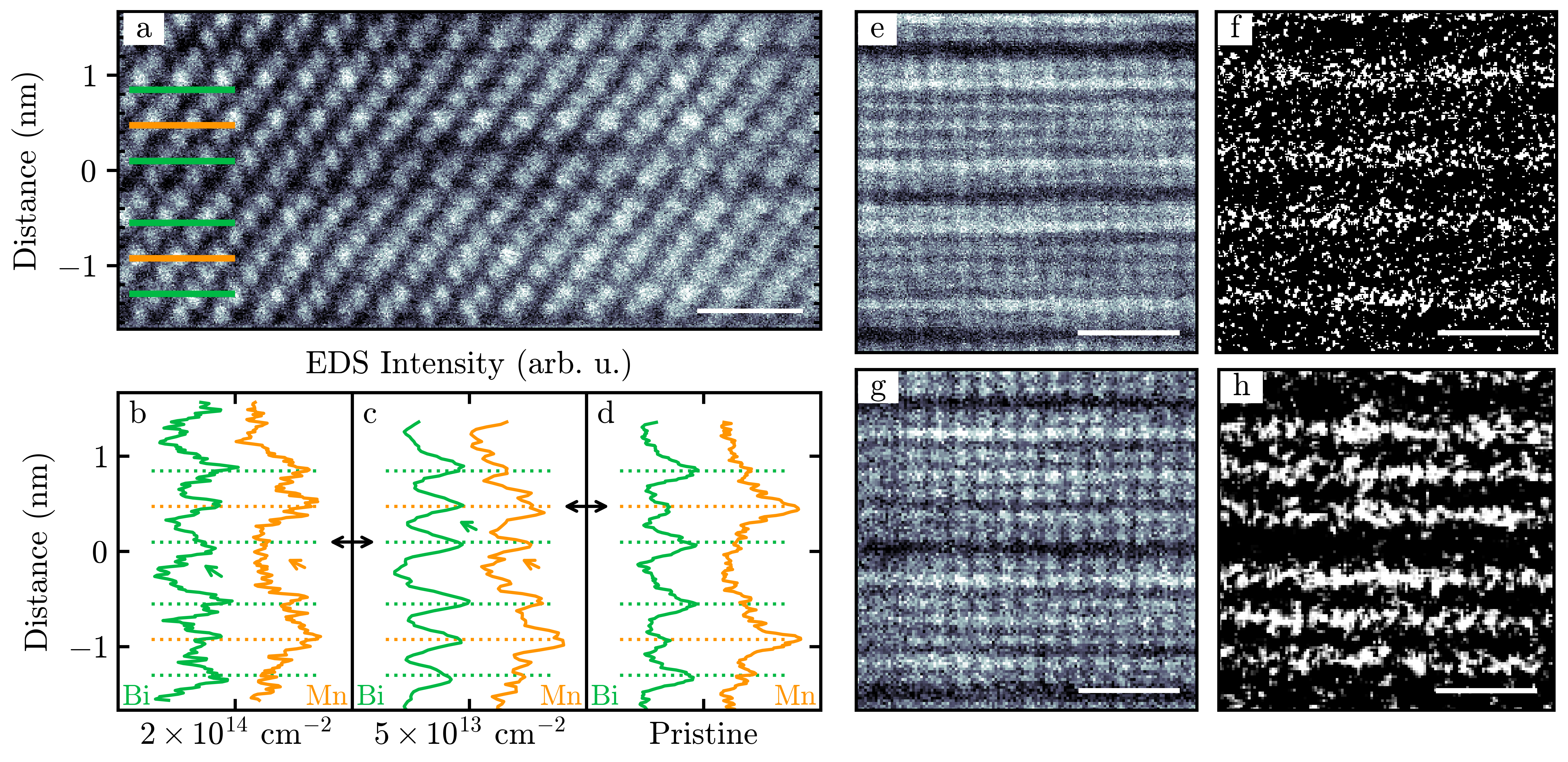}
    \caption{(\textbf{a}) STEM-image of a bilayer separating from a septuple layer after irradiation by a fluence of \fluence{2}{14}. (\textbf{b, c, d}) show EDS-intensities (intensity increasing from left to right) corresponding to the indicated fluence after integration along the horizontal direction. Dotted lines indicate cation planes which are also indicated on the STEM-image (\textbf{a}) as solid lines. Notable changes after irradiation are annotated, namely (\textbf{b}) the presence of Bi and negligible Mn in the detaching BL and (\textbf{c}) an increased cation intermixing after low fluence irradiation. Spatially resolved Bi M$_\alpha$ EDS maps and corresponding STEM micrographs are shown for the (\textbf{e}, \textbf{f}) pristine and (\textbf{g}, \textbf{h}) low fluence irradiated structures. All scale bars indicate a distance of 1 nm.}
    \label{fig:tem-edx}
\end{figure*}

Figure \ref{fig:tem-edx}(a) contains a micrograph of a ripplocation defect. Two SLs are present, the atomic planes in the bottom SL remain attached over the lateral range of the micrograph, the top SL contains the planar defect. Here, a bilayer detaches from the SL and occupies the vdW gap, locally forming a separate QL and BL. As annotated on the spatially integrated EDS signal in Fig. \ref{fig:tem-edx}(b) the Bi concentration is spread evenly across the QL and BL part of the ripplocation. In the region of the detaching BL, the spatially integrated Mn EDS curves decrease to background levels, while a clear presence is seen in the remainder of the SL. Remarkably, the distribution of Mn within the QL + BL structure is thus largely concentrated in the top five layers while scarce in the detaching BL. The QL layer part of the structure is therefore similar to that of Mn doped \bt. As has been noted in the context of surface collapse and reconstructions of MBT, Mn doped \bt\ is stable while Bi doped \mt\ is not~\cite{hou2020}. This chemical instability of Bi doped \mt\ is likely the reason for an absence of high Mn concentrations in the detached bilayers, even though intercalated \mt\ layers can be stabilized in MBT~\cite{chen2023}. The layers within the modified SL no longer align with the expected separation present in the near pristine SL below it, again emphasizing the adapted order in the $c$-axis. This is in contrast to the apparent alignment of the peaks of the integrated EDS signal to the pristine cation planes. The chemical disorder at this level of irradiation should therefore be interpreted as high, with significant intermixing over all lattice sites, resulting in a fine structure that is not clearly resolvable in the EDS signal.

The presence of predominantly Bi and Te within the detaching BL defects and the swapped intensity profile across the swapped bilayer leads us to conclude that the origin of planar defects generated by high fluence irradiation is the same as the source of swapped bilayer generation in \bt, \ce{Sb2Te3}, and \gst: cation-anion intermixing~\cite{li2025,wang2019,wang2021,deng2024,wang2018}. High concentration of cation-anion antisites render the SL structure unstable and a relaxation occurs to a new partially disordered structure. The key fact is the similar formation energies of both cation-anion antisite defects (here \bite\ and \tebi)~\cite{deng2024,li2025,wu2023,du2021}. Their coexistence breaks the normal crystal order and allows for the formation of complex planar defects. Such a mechanism is fundamentally different than, for example, the collapse of a structure due to a high vacancy concentration~\cite{hou2020}. We note that, although swapped bilayers and ripplocations are the dominant form of defects found after high fluence irradiation, other defect types are also present in lower abundance. These defects are either related to interstitials or are variations on the swapped bilayer and ripplocation defects (see Supplemental Material~\cite{SM}). We also note that some literature contains observations of singular swapped bilayer defects or ripplocations in MBT, but they are not necessarily recognized as such~\cite{hou2020,inoch2025,bai2024}.

In the low-fluence regime, the long-range crystallographic order of MBT is largely unaffected. As shown in the STEM micrographs of Fig. \ref{fig:tem-img}(b) and \ref{fig:tem-edx}(g) the periodic SL structure with a clear vdW gap remains after irradiation and is nearly identical to the pristine film. The loss of contrast between Bi and Mn planes does indicate substantial intermixing within the cation planes. This is clearly observed in the Bi M$_\alpha$ EDS maps in the pristine (Fig. \ref{fig:tem-edx}(f)) and low fluence irradiated films (Fig. \ref{fig:tem-edx}(h)). Where Bi primarily occupies the Bi planes in the pristine structure, after irradiation the Bi concentration is spread equally over the cation (Mn, Bi) planes. Some Bi intensity in the Mn plane is already present in the pristine film (\mnbi\ and \bimn\ antisites are the dominant native point defects found after growth~\cite{wu2023,lai2021,yuan2020,he2023,garnica2022, huang2020}), but the irradiation strongly increases it. Fitting of the spatially integrated EDS curves (Fig. \ref{fig:tem-edx}(b-d)) shows that \bimn\ antisites account for $(15\pm2)\%$ of the total Bi concentration, while this rises to $(31\pm1)\%$ after irradiation by the low fluence. The Bi can therefore be regarded as completely dispersed over all cation sites after irradiation conditions corresponding to a DPA (displacements per atom) of $0.17$ (estimated using SRIM). A similar analysis cannot be performed on the Mn concentration as the Mn EDS intensity is not sufficiently high. As annotated on the spatially integrated EDS curves (Fig. \ref{fig:tem-edx}(c)) there is however a clear presence of \mnbi\ antisite defects, as expected. In contrast to the heavily defected structure found after irradiation with a high fluence, the low fluence regime is characterized by cation site disorder, without significant changes to the global structure of the vdW material. This suggests that ion irradiation can serve as a viable method to controllably modify the cation antisite concentration and tune material properties through defect engineering~\cite{yan2022}. Similar to the high fluence case, other defects are found but to a lesser extent. Most notably, some cation-anion antisite defects can be observed (indicating the start of the transition to the high fluence regime) and Te interstitials within the vdW gap (see Supplemental Materials~\cite{SM}).

\subsection*{Self-Assembly of Cation-Anion Antisites}

The appearance of two-dimensional defects under ion bombardment is, at first sight, surprising, since irradiation directly produces point defects (vacancies, antisites, and interstitials) by displacing atoms from their lattice sites. Planar defect formation must therefore involve self-assembly of these point defects into extended structures. The intrinsically non-equilibrium nature of ion irradiation, in which energetic incident ions generate collision cascades and transient local heating that can promote dynamic annealing, renders a fully ab initio treatment impractical. We instead present density functional theory formation energies for point, line, and planar defects arising from Bi–Te intermixing, which reveal a systematic decrease in formation energy with increasing defect dimensionality, consistent with a point-defect-driven mechanism for planar defect nucleation.

\begin{figure}
    \includegraphics[width=\columnwidth]{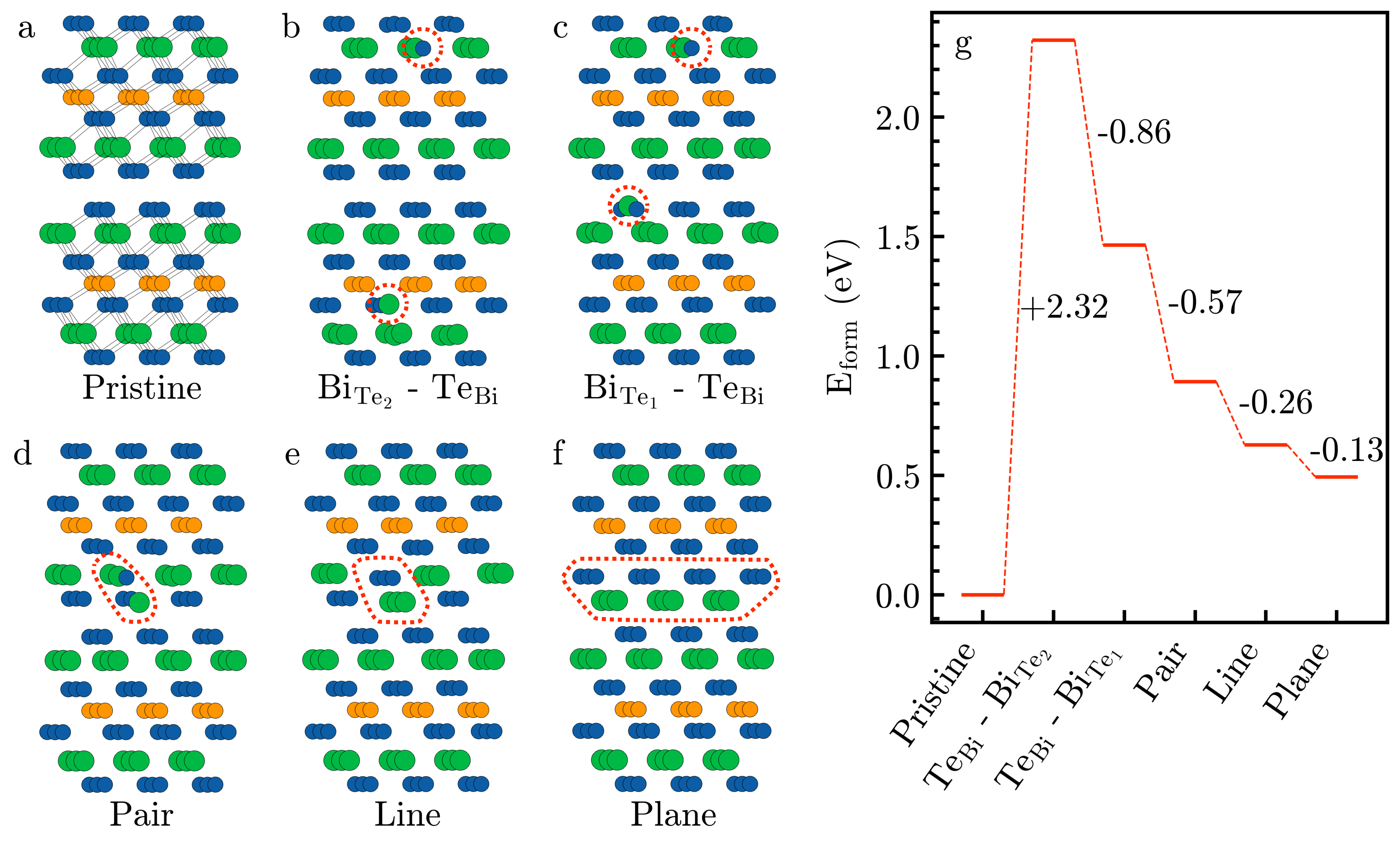}
    \subfloat{\label{fig:pristine}}    
    \subfloat{\label{fig:bite2}}       
    \subfloat{\label{fig:bite1}}       
    \subfloat{\label{fig:pair}}        
    \subfloat{\label{fig:line}}        
    \subfloat{\label{fig:plane}}       
    \subfloat{\label{fig:form_energy}} 
    \caption{Structural models and defect stability of Bi-Te defect configurations. (\textbf{a}–\textbf{f}) Side-view atomic models of the studied systems: (\textbf{a}) pristine structure, (\textbf{b}) \bitetwo\ antisite defect, (\textbf{c}) \biteone\ antisite defect, (\textbf{d}) defect pair, (\textbf{e}) linear defect cluster, and (\textbf{f}) planar defect arrangement. Red dashed circles (panels \textbf{b}, \textbf{c}) and areas (panels \textbf{d}–\textbf{f}) highlight the local structural modifications of the defects. (\textbf{g}) Calculated formation energy ($E_{\text{form}}$) in units of eV per pair for each configuration. The pristine system is used as the energy reference ($0$ eV). Dashed lines and numerical labels indicate the energy differences between successive defect states, illustrating the energy stabilization as defects aggregate from isolated sites into swapped planar structures.}
    \label{fig:dft} 
\end{figure}

To elucidate this thermodynamic pathway underlying the structural reconfiguration from uncoordinated antisites to planar defects, six distinct configurations were modeled (\cref{fig:pristine,fig:bite2,fig:bite1,fig:pair,fig:line,fig:plane}): the pristine structure, isolated Bi$_{\text{Te}}$ antisite defects in both Te$_1$ and Te$_2$ sites, an aggregated defect pair, a defective line, and a defective plane. To retain material stoichiometry, a \tebi\ antisite is introduced alongside the isolated \bite\ antisites, placed in a separate vdW layer to prevent interactions between the antisite defects.

Since all configurations are stoichiometric, the formation energy per defective pair ($E_{\text{form}}$) is defined as:
\begin{equation*}
    E_{\text{form}} = \frac{E_{\text{defective}} - E_{\text{pristine}}}{n_{\text{pairs}}}
\end{equation*}
where $n_{\text{pairs}}$ represents the number of aggregated defect pairs. The structures corresponding to defect pairs, defect lines, and a defect plane contain one, three, and nine pairs of agglomerated antisite defects, respectively.

\Cref{fig:form_energy} illustrates the results of the formation energy, revealing that creating a pair of isolated defects requires 2.32 eV/pair for \bitetwo-\tebi\ and 1.46 eV/pair for the \biteone\ site. \biteone\ is favored by 0.86 eV/pair, which aligns with the experimental observations of defect agglomeration near the vdW gap (in the terminating bilayer). Furthermore, our results are in agreement with previous calculations, which predict a higher formation energy for the \bitetwo\ antisite than for the \biteone\ antisite~\cite{wu2023}. Since it is energetically favored, and consistent with the experimental observations, we fix one \biteone\ antisite in the terminating anion layer and investigate the formation energy of defect structures containing this antisite with increasing dimensionality. An energy gain from clustering is observed, similar to that observed for $\mathrm{Bi_{2}Te_{3}}$~\cite{li2025}. Upon agglomeration into a pair, the energy per pair decreases by 0.57 eV. If a defective line forms, the energy is further reduced by 0.26 eV/pair, and the plane decreases the system energy by 0.13 eV/pair. \del{The monotonic decrease of the formation energy with increasing dimensionality indicates that this agglomeration process is spontaneous once ion bombardment generates the Bi-Te antisites and provides the initial kinetic energy for atomic displacement.} \add{The monotonic decrease of the formation energy with increasing dimensionality indicates that this agglomeration process is energetically favorable. The formation of planar defects may be spontaneous, once ion bombardment generates the Bi-Te antisites and provides the initial kinetic energy to overcome kinetic barriers to antisite migration and clustering.}

To investigate the structural effects of bilayer defects, large-scale modeling beyond the capabilities of standard DFT methods is required. Consequently, we used the \texttt{MACE-medium-MPA-0} model~\cite{Batatia2025}, trained on the Materials Project and Alexandria datasets, to perform a large-scale simulation. To capture long-range vdW interactions, we augmented the potential energy of the model by explicitly summing the DFT-D3 correction term. We constructed a supercell containing 1,120 atoms, featuring three distinct regions: pristine, partially intermixed with Bi-Te antisite defects, and fully swapped. The configuration was designed to maintain the periodic boundary conditions across the lateral interfaces. The structure was subsequently relaxed using the Atomic Simulation Environment (ASE). 

\begin{figure*}
    \includegraphics[width=\textwidth]{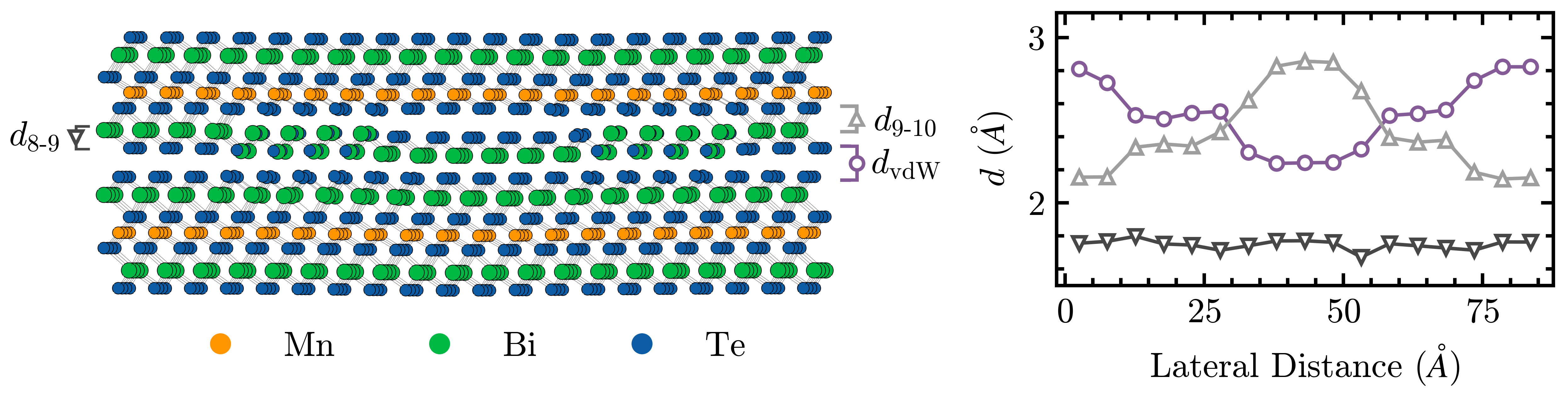}
    \subfloat{\label{fig:largemodel}}    
    \subfloat{\label{fig:spacing}}       
    \caption{Structural relaxation of Bi-Te bilayer defects. (\textbf{a}) Relaxed supercell displaying pristine, partially intermixed, and fully bilayer swapped regions. Mn, Te, and Bi atoms are shown in orange, blue, and green, respectively. (\textbf{b}) Interlayer spacing $d$ across the lateral distance of the relaxed structure. A contraction of the vdW gap ($d_{\text{vdW}}$) is compensated by the expansion of the $d_{9-10}$ spacing. The separation within the detaching bilayer ($d_{8-9}$) remains constant throughout the swap, consistent with HAADF-STEM data.}
\end{figure*}

\Cref{fig:largemodel} shows the relaxed supercell, while \cref{fig:spacing} illustrates the local interlayer spacings as a function of the $x$-coordinate (see Supplemental Material for details~\cite{SM}). Our results reveal a progressive contraction of the vdW gap, compensated by an expansion between layers 9 and 10. As a result, the separation within the detaching bilayer ($d_{8-9}$) remains nearly constant after the swap, a result consistent with the structural HAADF-STEM data shown in Fig. \ref{fig:tem-img} and \ref{fig:tem-edx}.

A simple heuristic picture of the self-assembly also emerges: \bite\ antisites act as acceptors, whereas \tebi\ act as donors~\cite{du2021, he2023}. As in classic donor-acceptor defect pairing (for example, nitrogen-vacancy pairs in diamond and silicon-vacancy pairs in GaAs), the uncompensated defect charge is likely a good predictor of defect dynamics in MBT. The subsequent agglomeration into planar defects is anticipated to be specific to vdW materials, where the absence of a rigid three-dimensional framework allows complex structures to form near the vdW gap.

\subsection*{Macroscopic Detection of the Disordered Phase}

\begin{figure}
    \includegraphics[width=\columnwidth]{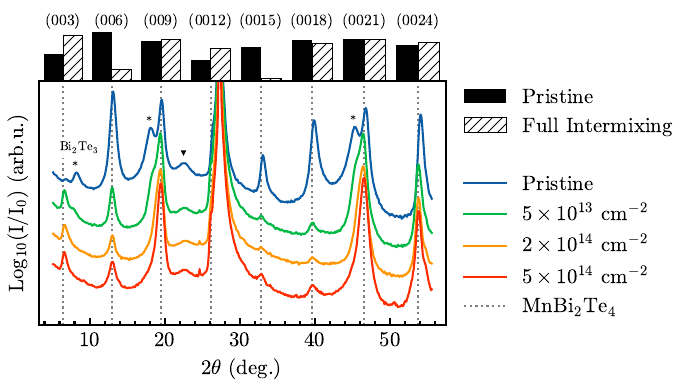}
    \caption{Diffractograms measured on pristine and irradiated films. Dotted lines indicate expected diffraction peaks for the pristine \mbt\ structure. The asterisks indicate some \bt\ diffraction peaks that are notable in the diffractograms. The triangle indicates a diffraction peak from the capping (see Methods). Top: Simulated XRD intensities shown in log-scale for the pristine and fully intermixed structure, as discussed in the main text.}
    \label{fig:xrd}
\end{figure}

The X-ray Diffraction (XRD) experiments (shown in Figure \ref{fig:xrd}) reveal that ion irradiation of \mbt\ causes a systematic change in the diffraction intensities in the $(00L)$ series. A significant increase in the $(003)$ diffraction (the reciprocal of the SL separation) and decrease in the $L=\{6,\ 15,\ 18\}$ peaks is observed. For vdW films with layer disorder, the Bragg angles and respective diffraction intensities are strongly affected \add{by the interference between X-rays scattered from the individual vdW layers, which depends on their order and spacing}\del{by X-ray rescattering and other phenomena related to dynamical diffraction theory}~\cite{kagerer2020,morelhao2017,morelhao2019, penacchio2022, mallick2026}. \add{Theoretical frameworks to calculate diffraction intensity in these systems exist, but assume perfectly flat atomic planes. The microscopic observation of planar defects at high concentration after irradiation therefore restricts the usage of these models.} We thus refrain from making quantitative statements about the diffractograms. We note however that simple modeling based on static diffraction intensity of a completely intermixed crystal structure (i.e. complete chemical Mn-Bi-Te intermixing) qualitatively reproduces the change in diffraction intensities (as shown in the top of Fig. \ref{fig:xrd}). The predicted increase and decrease of intensity as compared to the pristine structure follows the same trends observed in the irradiated films (besides the $(0018)$ diffraction peak). 

Shifts in the diffraction peaks are also evident, most prominently the \bt\ peaks (originating from intercalated layers and the buffer film). \add{After low fluence irradiation STEM measurements (see large scale micrograph in Supplemental Material~\cite{SM}) show that the intercalated BT layers remain present.} With further increasing fluence, the whole film evolves to a single, more homogenized, structure with identical diffraction features.  The BT buffer layer experiences significant intermixing with the adjacent part of the MBT film; the evolution of the BT diffraction peaks should therefore be interpreted with caution. It is however known that swapped bilayers and ripplocations can be stabilized in BT in high concentrations~\cite{deng2024}. With the mechanism of cation-anion intermixing in mind, the evolution of the MBT film and BT buffer to the same structure may hint at a universal disordered phase in the vdW layered chalcogenides attainable by ion irradiation. Even at a higher fluence of \fluence{5}{14}, the system shows periodic order: the partially disordered phase persists even up to a displacement per atom of 1.65, without reaching amorphisation. \del{In addition to confirming the retention of periodic order along the $c$-axis, XRD thus offers a practical means to screen and monitor fluence regimes in the \fivesix\ family and vdW materials more generally.}

\subsection*{Magnetism and Transport}

\begin{figure}
    \includegraphics[width=\columnwidth]{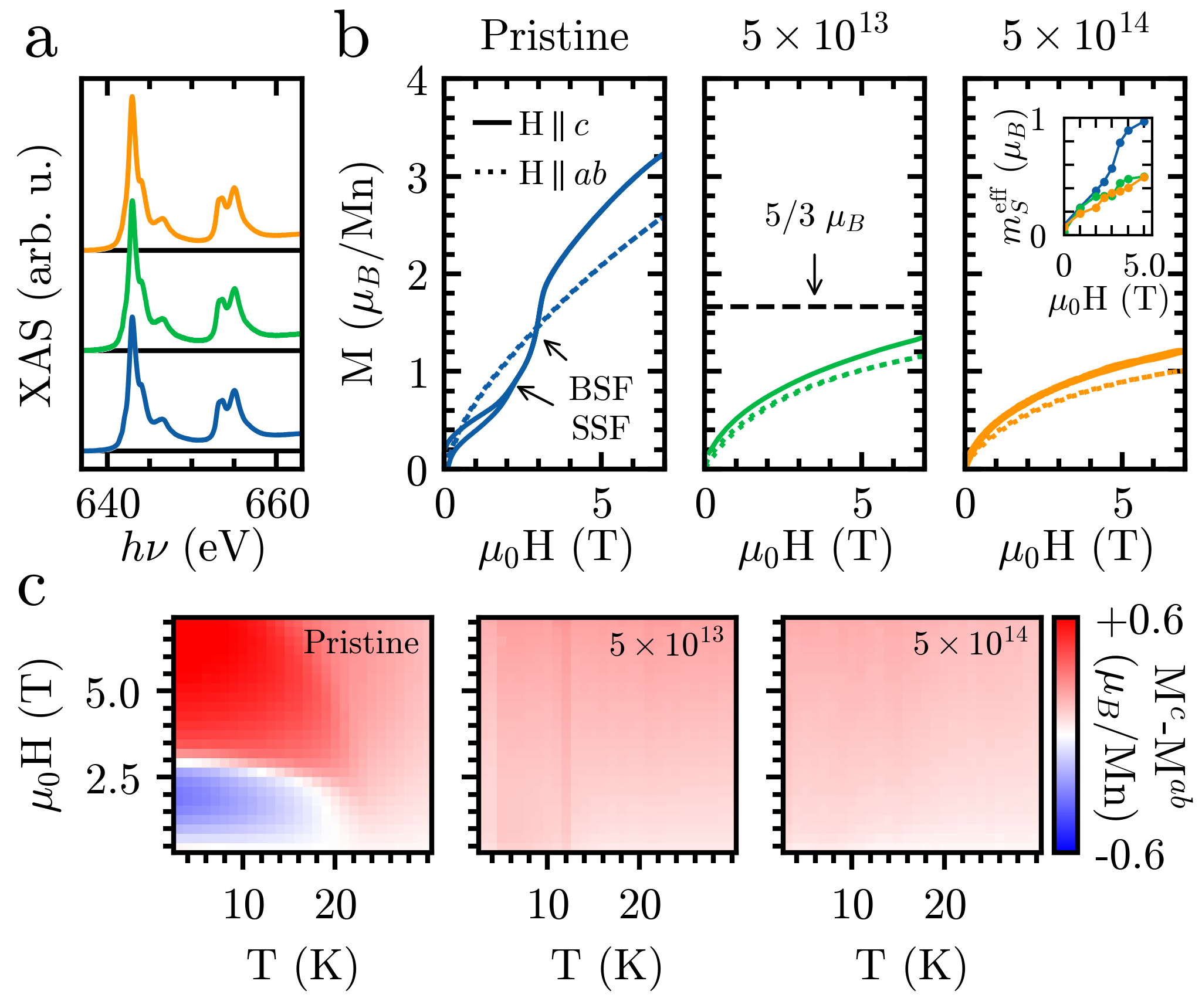}
    \caption{\add{(\textbf{a}) X-ray Absorption Spectra of the Mn $L_2$ and $L_3$-edge measured as a function of photon energy $h\nu$ at normal incidence in total fluorescence yield for each fluence at 2 K. 
    (\textbf{b}) Magnetization $M$ as a function of applied magnetic field $\mu_0H$ at 2 K for out-of-plane ($H\parallel c$) and in-plane ($H\parallel ab$) fields measured by vibrating sample magnetometry. Fluences are indicated on top of the figure in cm$^{-2}$. Inset figure on the right shows effective spin moment $m_S^{\text{eff}}$ from XMCD under normal incidence and ($H\parallel c$) at 2 K.
    (\textbf{c}) Temperature and field dependence of the difference between out-of-plane and in-plane magnetization. Raw $M(H)$ curves are provided in the Supplemental Material \cite{SM}.}}
    \label{fig:MH}
\end{figure}

To evaluate how ion irradiation can be used as a means to control the behavior of MBT, we systematically probe its magnetic and transport response.
Figure \ref{fig:MH}(a) shows X-ray absorption spectroscopy (XAS) of the Mn atoms in the pristine and irradiated films. The spectra are identical for all levels of disorder. The shape of the spectrum is in line with reported XAS and shows the Mn population is in a (high spin) \threedfive\ state~\cite{otrokov2019, kagerer2020, zeugner2019, hirahara2020}. Changes in the magnetic and transport properties of the film can therefore be attributed to the chemical and structural disorder induced by ion irradiation, rather than to a change in the Mn electronic state. \del{The observed modifications reflect the cumulative effect of irradiation-induced defects, without requiring a significant alteration of the local Mn environment.}

Pristine \mbt\ possesses an A-type antiferromagnetic (AFM) order that is well modeled by an AFM chain with uniaxial magnetic anisotropy~\cite{lai2021}. Surface and bulk-spin-flop transitions (SSF \& BSF resp.) occur under moderate magnetic field along the $c$-axis. Magnetic field values of the SSF and BSF match literature values, as previously reported~\cite{bac2022}. Although saturation is not reached, the measured moment is consistent with a saturation magnetization of $\sim 4\ \mu_B/\text{Mn}$~\cite{yan2019, lai2021}. A minor hysteresis is present at low field values which has been discussed in literature~\cite{bac2022,yang2021,su2021,lian2025}. \add{Magnetic fields applied perpendicular to the easy axis ($H\parallel ab$) induce continuous canting of the spins until saturation. This canting at low field does not occur for out-of-plane fields, as the magnetic anisotropy favors a collinear alignment of the spins along the $c$-axis. The in-plane magnetization ($M^{ab}$) is therefore greater than the out-of-plane magnetization ($M^{c}$) before the spin-flop transition. At field strengths greater than the spin-flop field the Zeeman energy overcomes the exchange and anisotropy energy and forces the spins into a canted non-collinear order \cite{bogdanov2007}, resulting in $M^{c}>M^{ab}$ until saturation.}

\add{After ion irradiation, the field driven spin-flop transitions are not observed and the magnitude of the magnetization is halved at high field (Fig. \ref{fig:MH}(b)).} X-ray Magnetic Circular Dichroism (XMCD) experiments provide the same insight (see Methods). \add{The spins of \mnbi\ antisites align antiparallel with Mn$_{\text{Mn}}$ spins in \mbt\ at low temperature \cite{lai2021}. Under complete cation intermixing in an otherwise pristine lattice structure, this alignment produces an upper limit on the magnetization of $5/3\ \mu_B$ per Mn atom. As annotated on Fig. \ref{fig:MH}(b), the magnetization of the system after irradiation does not cross this threshold.} This behavior is consistent with the expected magnetic interactions in the system: despite the redistribution of Mn atoms, including onto Bi sites, the overall Mn concentration remains high, such that antiferromagnetic superexchange mediated by the anion sublattice continues to dominate~\cite{goodenough1976,kanamori1959,anderson1959}. 

\add{The directional field dependence of the magnetization in the as-prepared film shows the signature features of an AFM order with uniaxial anisotropy. After irradiation, however, the difference between $M^{c}$ and $M^{ab}$ strongly decreases. Over the whole field range, $M^{ab}$ is smaller than $M^{c}$, by an amount that can be attributed to size effects during volumetric magnetometry measurements. Figure \ref{fig:MH}(c) shows this difference in $M^{c}$ and $M^{ab}$ as a function of applied field strength as measured up to a temperature of 30 K. Below the N\'eel temperature, the as-prepared film retains the behavior observed at low temperatures: $M^{ab}$ is larger than $M^{c}$ before the BSF transition, and after the BSF transition $M^{c}>M^{ab}$ until saturation. After irradiation, no change in $M^{c}-M^{ab}$ is observed in the same temperature range. These measurements show that the macroscopic (as-measured) magnetic anisotropy of \mbt\ is reduced after irradiation. This reduction is not synonymous with a loss of magnetic anisotropy on the atomic scale, but likely reflects a disordering of the anisotropy tensor (as further discussed below).}

\begin{figure}
    \includegraphics[width=\columnwidth]{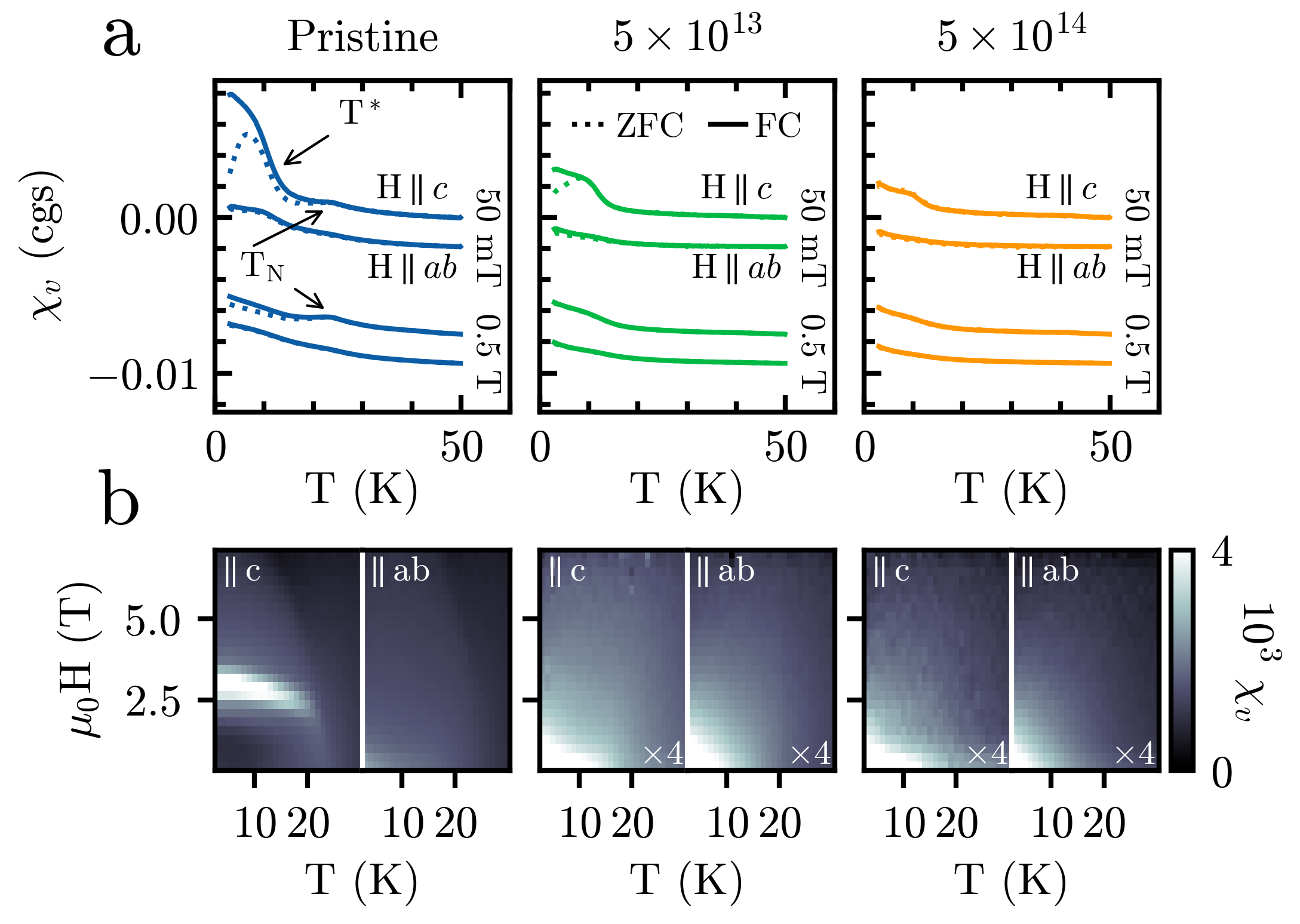}
    \caption{\add{(a) Temperature dependence of the magnetization measured in both out-of-plane ($H \parallel c$) and in-plane ($H \parallel ab$) directions for the irradiation series. The volumetric susceptibility $\chi_v$ is calculated as $M/H$ in cgs units. Zero-field-cooled (ZFC) and field-cooled (FC) curves are given as dotted and solid lines, respectively. The applied field strength $\mu_0H$ is annotated on the right-hand-side. Curves are vertically offset for readability. (b) Field dependence of $\chi_v = \text{d}M/\text{d}H$ for each fluence (as indicated on top in cm$^{-2}$) in out-of-plane (left) and in-plane directions (right). Raw curves are provided in the Supplemental Material \cite{SM}. The magnitude of $\chi_v$ is multiplied by four for the irradiated films for readability.}}
    \label{fig:MT}
\end{figure}

\add{Figure \ref{fig:MT}(a) shows that the N\'eel transition $T_{\text{N}}$ of the as-prepared film is observed as a cusp with a maximal susceptibility at 22.6(1) K (temperature step taken as the uncertainty). In an out-of-plane field configuration, a small difference between zero-field-cooled (ZFC) and field-cooled (FC) curves is observed below $T_{\text{N}}$, consistent with uncompensated moments in the interlayer AFM order. The uniaxial anisotropy in \mbt\ results in a temperature-independent susceptibility below $T_{\text{N}}$ under a constant in-plane field. As a result, the transition is hardly distinguishable from the background when $H\parallel ab$. After irradiation, for both fluences, no magnetic transition is observed in the expected temperature range for pristine \mbt. The absence of this transition in $M{(\text{T})}$ data does not in itself prove the absence of an interlayer AFM order. As shown in the as-prepared film, the experimental signature of the N\'eel transition can be difficult to observe, while interlayer AFM is present. To compensate for this shortcoming of $M(T)$-curves, we measured the detailed field dependence of the system at a series of fixed temperatures. As shown in Figure \ref{fig:MT}(b), the as-prepared film shows clear field-driven transitions along and perpendicular to the easy axis up to $T_{\text{N}}$. Notably, the spin-flop transitions are observed for $H\parallel c$, and saturation for both $H\parallel c$ and $\parallel ab$. After irradiation, this is not the case, and only a low-field nonlinearity in the $M(H)$ curves remains. There is no clear saturation of the spins at high field after ion irradiation, even close to the transition temperature. We therefore consider survival of the interlayer AFM order after ion irradiation by a fluence equal to or greater than \fluence{5}{13} unlikely.}

\add{In all films a low temperature magnetic transition is present, most clearly observed under an out-of-plane field of 50 mT. A more expansive treatment is given in the Supplemental Material~\cite{SM}. In the pristine structure, this transition can be linked to the freezing temperature $T^*$ of \mnbi\ antisite spins in \mbt\ \cite{sahoo2024}. After ion irradiation, an exact statement on the nature of the observed transition cannot be made, but basic characterization is possible. The susceptibility of both irradiated films is anisotropic in field direction. Moreover, after irradiation with the low fluence, under a 50 mT field, the bifurcation of ZFC and FC curves is greater for out-of-plane than for in-plane fields. This is no longer true after irradiation in the high fluence regime, where the anisotropy in bifurcation magnitude is not present. The $M(T)$ curves suggest that, at atomic scale, there are energetically preferred directions, which affect the pinning and energetic barriers separating spin configurations upon cooling.}

\add{The loss of macroscopic anisotropy in the field-dependent magnetization, combined with persistent low-field anisotropy in $M(T)$-curves, is consistent with local disorder in the easy-axis direction while the anisotropy energy itself remains non-zero. At low fields, the Zeeman energy is in competition with the anisotropy energy, which is locally able to create domains or pinned spin directions. Once the Zeeman energy exceeds the local anisotropy energy, the moments are pulled away from their microscopic easy-axes and progressively aligned with the field, so that the approach to saturation becomes largely independent of the crystallographic orientation of the film. For each Mn atom, the easy axis direction is determined by its crystallographic environment. Ion irradiation produces point (and planar) defects that can lower the site symmetry. Moreover, the local configuration of these defects determines the atomic environment of the Mn atoms, which therefore varies from one Mn site to the next. The loss of macroscopic anisotropy is therefore consistent with a degradation of the local easy-axis directions through chemical (and structural) disorder, while the anisotropy energy itself remains non-zero. These arguments extend to the film irradiated in the high fluence regime, where the absent anisotropy in the bifurcation magnitude may reflect further disordering of the local easy-axes, a reduced anisotropy energy, or both.}

\add{Finally, we note that the magnitude of the magnetization after ion irradiation is consistent with an AFM intralayer exchange between Mn spins. The observed low temperature transition may be related to these interactions. The transition temperature does not vary greatly as disorder increases (see Supplemental Material~\cite{SM}), which indicates that the energy scale of the spin Hamiltonian is not significantly modified. The retention of intralayer AFM exchange while interlayer AFM order is lost is likewise consistent. A reduction of the effective moment per vdW layer, by the creation of new Mn antisites whose spins antialign with each other, decreases the interlayer exchange energy, even if the interlayer exchange coupling is not affected by ion irradiation.}

\begin{figure}
    \includegraphics[width=\columnwidth]{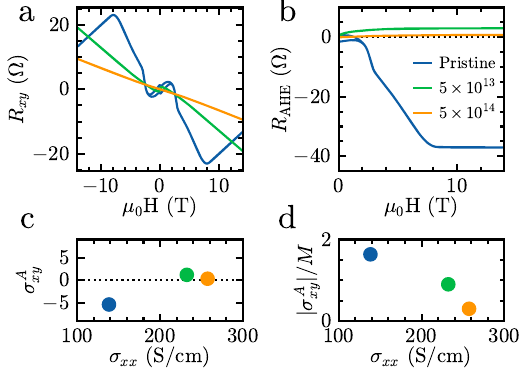}
    \caption{\add{(a) Transverse resistance $R_{xy}$ measured as a function of out-of-plane magnetic field $\mu_0H$ at 2.0 K. Contact convention relates a negative Hall slope to electron-dominated transport. (b) Anomalous Hall resistance \rahe\ obtained by subtracting the ordinary Hall effect from $R_{xy}$ at high field. (c) The anomalous Hall \sahe\ and longitudinal \sxx\ conductivity values extracted at 14 T (uncertainty smaller than marker size). (d) $|\sigma_{xy}^A|$ (14 T) normalized by $M$ (7 T, below saturation; ratio is thus a conservative upper bound) versus \sxx. All conductivity values are in units S/cm and magnetization $M$ is in units $\mu_B$/Mn.}}
    \label{fig:trans}
\end{figure}

\add{Next, we examine the impact of irradiation on the anomalous Hall response of MBT.} The Berry curvature, together with the magnetization, governs the magnitude of the anomalous Hall effect (AHE) in MBT \cite{bac2025}. \add{As shown in Fig. \ref{fig:trans}(a) the Hall resistance $R_{xy}$ contains both a linear ordinary Hall effect (OHE) at high field, and an anomalous Hall response at low field. The anomalous Hall resistance (\rahe)} extracted by subtracting a linear slope at high fields is plotted in Fig. \ref{fig:trans}(b). \rahe\ decreases sharply from the pristine to the low-fluence irradiated film. In the planar disordered phase reached at high fluence, \rahe\ is further suppressed and becomes nearly negligible. The longitudinal conductivity (\sxx) is enhanced with increasing fluence, while the magnitude of the anomalous Hall conductivity (\sahe) decreases \add{fourfold after low-fluence irradiation ($-5.3(2)$ $\rightarrow$ $1.23(3)$ S/cm), and an order of magnitude after irradiation in the high fluence regime ($-5.3(2)$ $\rightarrow$ $0.37(1)$ S/cm), see Fig.~\ref{fig:trans}(c)}. The suppression of \sahe\ \add{is} greater than the decrease in magnetization\add{, after irradiation the magnetization only halves (Fig. \ref{fig:MH}(b) and \ref{fig:trans}(d)). This points to a direct modification of the mechanism underlying the AHE in the system. Irradiation likewise causes a change in the sign of \sahe, an effect that cannot be attributed to simple disorder broadening.} In the low fluence regime, where the irradiation induced disorder is dominated by cation antisites, the degradation of \add{$|\sigma_{xy}^A|$ is in line with a suppression of Berry curvature known to cause the anomalous Hall effect in pristine MBT. A suppression of Berry curvature is also consistent with the predicted weakening} of band inversion in \add{\mbt}\ under high levels of cation intermixing~\add{\cite{chen2026}}. \add{The change in the sign of \sahe\ is consistent with density functional theory calculations on pristine \mbt, which indicate that the sign of the Berry curvature contribution varies over the band structure~\cite{chen2025}. Contributions from extrinsic mechanisms (side-jumps) in the irradiated samples cannot be excluded by these experiments \cite{nagaosa2010}. However, it is difficult to reconcile an extrinsic origin responsible for the pronounced suppression and sign reversal of \sahe\ without a modification of the intrinsic contribution. Our observations are therefore most consistent with an irradiation-induced suppression of the occupied states' Berry curvature as the dominant cause of the reduced anomalous Hall response.} 

Beyond their effect on magnetic and \del{topological behavior} \add{anomalous Hall response,} the observed structural changes are directly reflected in the electrical transport properties. \add{As shown in Fig. \ref{fig:trans}, the slope of the OHE at high field varies between the pristine and irradiated films in both magnitude and sign. Our MBT films are grown on a Bi$_2$Te$_3$ buffer layer, however, as further discussed in the Supplemental Material~\cite{SM}, the change in sign of the Hall slope can not be attributed to the buffer layer alone, and requires a change of majority carrier type in \mbt.} The as-prepared film exhibits $p$-type majority carriers\add{, with a sheet carrier density $|n| = 3.6\times10^{14}$ cm$^{-2}$ extracted from the OHE (see Supplemental Material for details~\cite{SM}). The carrier type is} consistent with MBE-grown MBT where \mnbi\ acceptor defects are intrinsically present after growth~\cite{he2023}. Following low-fluence irradiation, the \add{majority} carrier type switches to $n$-type \add{($|n| = 3.9\times10^{14}$ cm$^{-2}$)}. As discussed above, this regime is characterized by a nearly complete redistribution of Bi over all cation sites. Given that \bimn\ antisites act as donor defects~\cite{wu2023, lai2021, yuan2020, he2023, garnica2022, huang2020} and become dominant due to the material stoichiometry, this cation intermixing naturally explains the observed transition to $n$-type behavior. The \add{majority} carrier type remains $n$-type in the high-fluence regime \add{($|n| = 8.6\times10^{14}$ cm$^{-2}$)}. In the planar disordered crystal, \add{the further increase in carrier density may be related to new carriers generated by the bond disruption required to form swapped bilayers and ripplocations. Further information regarding data treatment, carrier density, mobility, and the temperature dependence of the resistance is provided in the Supplemental Material~\cite{SM}.} \del{the increase of \sxx\ combined with the retention of a similar \sahe\ as compared to the low fluence irradiated film suggests the generation of new (trivial) carriers that do not contribute to transverse transport. The swapped bilayers and ripplocations characteristic of this regime are associated with local bond disruption within the vdW layer, which may provide a structural pathway for the introduction of such carriers.}

\section*{Discussion}

One of the central, underlying outcomes of this work is the identification of a novel disordered phase in \mbt\ induced by high-fluence irradiation. On its own, this carries important implications for the understanding of defect formation and accumulation in vdW materials under high-energy particle irradiation. Remarkably, the system resists amorphisation up to a displacement per atom of at least 1.65 (the maximum reached in the present study). This is well above typical amorphisation threshold of covalent semiconductors such as Si and Ge, and III–V semiconductors such as GaAs and InP. This radiation resistance arises from the ability of irradiation-induced point defects to reorganize into complex microstructures that retain partial crystallographic order. This mechanism is fundamentally different from that observed in classical irradiation-resistant materials, such as \ce{ZnO} and \ce{CdTe} in the \ce{II-VI} semiconductor family~\cite{wesch2012}, governed by efficient dynamic annealing that suppresses defect accumulation. In contrast, in MBT, the periodic order is to a large extent preserved because the damage is accommodated by the coexistence and self-organization of multiple defect types. This behavior can be related to a general framework proposed for complex oxides, where materials that can energetically accommodate disorder (through mechanisms such as cation antisites and anion Frenkel defects) exhibit enhanced resistance to amorphisation by transforming toward disordered crystalline states rather than accumulating damage~\cite{sickafus2007}. Nevertheless, the self-assembly into extended planar defects along a well-defined lattice direction (characteristic of the layered vdW structure) is a distinctive feature of vdW materials, unraveled in the present work. Notably, the same coexistence of low-energy point defects that stabilizes the irradiated phase is also responsible for the narrow and challenging growth window of MBT~\cite{wu2023}. The results presented here therefore motivate further investigation into the hypothesis that vdW materials that are challenging to stabilize during (non-)equilibrium growth may, for that same reason, display enhanced resistance to ion irradiation and implantation.

\del{Another interesting aspect of this work is that the signatures of the transition from the pristine to the layer-disordered structure can be inferred from XRD measurements combined with basic structural modeling. However,} \add{The principal notion that supports this work is} the direct identification and detailed understanding of the underlying defects, such as swapped bilayers and ripplocations, which requires real-space imaging by STEM, where these features are unambiguously resolved. \add{Although X-ray diffraction provides macroscopic proof that periodicity remains after ion bombardment, the existence of widespread planar disorder that locally breaks translational symmetry along the $c$-axis cannot simply be inferred from it.}
This has important implications for the broader characterization of vdW materials. Standard approaches often rely on macroscopic techniques or surface-sensitive probes, which may not explicitly capture the presence of extended planar defects or defect self-assembly. \add{Here, without prior knowledge of planar defects, we show the X-ray diffraction experiments can be interpreted as consistent with substitutional disorder alone.} As a result, such defects can remain overlooked or be indirectly misinterpreted if not specifically considered in the analysis. The identification of cation–anion intermixing as the driving mechanism for the formation of swapped bilayers and ripplocations further suggests that similar defect structures may arise in other materials with comparable antisite energetics, such as the \ce{(V)_2(VI)_3} family. We therefore propose that layer-disordered phases, vdW-specific planar defects, and their self-assembly are likely more widespread than currently recognized.

Finally, we consider the broader implications of these findings for defect engineering in MBT. In the low-fluence regime, ion irradiation generates cation antisite defects, up to a nearly complete redistribution of Bi over all cation sites, without disrupting the long-range crystallographic order. This establishes a regime in which ion beams can be used to tune material properties through controlled defect engineering. The evolution of the Fermi level with increasing disorder is likely non-monotonic. In addition to cation antisites, irradiation introduces cation–anion antisites and vacancies at lower concentrations, leading to a more complex doping landscape when \bimn\ donor defects do not dominate the defect population. At the same time, the observation that the dominant irradiation-induced defects correspond to the most common native growth defects suggests a more universal and predictable response to ion irradiation within the TM-doped \ce{(V)_2(VI)_3} family than so far recognized. Beyond its conventional role in Fermi level tuning, we demonstrate that ion irradiation also enables control over \add{the magnitude and sign of $\sigma_{xy}^A$, in line with a modification of the carriers'} Berry curvature. \del{In this way,} \add{This suggests that} the topological phase of the system can be tuned while preserving periodic order, opening a route toward functionalization of this class of materials. Combined with masked irradiation, this approach \add{may enable} \del{enables} patterning of topological phases, enabling the design of lateral heterostructures.

\section*{Conclusion}

In summary, we demonstrate that ion irradiation enables controlled access to distinct defect regimes in MnBi$_2$Te$_4$, spanning from cation-disordered but structurally intact films to a layer-disordered phase dominated by van der Waals–specific planar defects. The emergence of this disordered phase highlights a mechanism in which irradiation-induced point defects reorganize into extended defect structures that preserve partial crystallinity, providing an alternative route to radiation tolerance in layered materials. \del{Importantly, the persistence of antiferromagnetic order alongside a strong reduction in magnetic anisotropy underscores the robustness of superexchange interactions even in highly disordered environments.}

Beyond structural and magnetic effects, we show that defect engineering through ion irradiation directly impacts the \del{topological properties of} \add{mechanism underlying the anomalous Hall effect in} the system. The strong suppression \add{and sign change} of the anomalous Hall response, disproportionate to the change in magnetization, \add{are consistent with the Berry curvature being an ion irradiation tunable parameter}\del{identifies Berry curvature as a tunable parameter governed by disorder and symmetry breaking}. \del{This establishes a framework in which} \add{This highlights the potential that} ion beams can be used not only to modify carrier density and defect populations, but also to tailor topological phases.

More broadly, these findings suggest that van der Waals materials with competing defect configurations may exhibit intrinsic resilience to irradiation through defect self-organization, and that such behavior may be more widespread across the tetradymite family than currently recognized. \add{The identification of a fundamental link between defect self-assembly and radiation tolerance motivates further work on the dynamical aspects of beam-solid interactions in this material class, which may be accessible through control of temperature and ion flux.} The ability to combine ion irradiation with lithographic masking further opens the perspective of spatially patterning topological phases, offering new opportunities for device architectures based on lateral heterostructures of magnetic topological materials.

\section*{Methods}

\subsection*{Material Synthesis \& Ion Irradiation}
MnBi$_2$Te$_4$ films are synthesized by MBE on GaAs(111)B substrates. The substrates are initially annealed up to 580$^{\circ}$C to desorb the native surface oxide. The GaAs surface
is then treated with a Te flux at 600$^{\circ}$C for 20 min to obtain a Te-termination. This step is critical to obtain a flat interface, and a smooth layer. A Bi$_2$Te$_3$ buffer layer (10 quintuple layers) is then grown at 340$^{\circ}$C. The Bi$_2$Te$_3$ layer is further annealed at 400$^{\circ}$C under a Te flux to further improve surface smoothness. To grow MnBi$_2$Te$_4$, we then sequentially exposed the substrate to a flux of the following: Mn-Bi-Te (for 30 s), Mn-Te (30 s) and Te (180 s). This is repeated 40 times, all while maintaining a substrate temperature of 320$^{\circ}$C. The growth is carried out under Te rich conditions. After the growth, the film is cooled down to room temperature and is covered with a capping layer of amorphous Te ($\sim$100 nm) and amorphous Se (10 nm) to protect from oxidation.

To avoid energy losses during irradiation in the thick capping and to allow fluorescent XAS measurements the capping is later removed by thermal annealing at 200 \textdegree C in an ultra-high-vacuum chamber with a $10^{-10}$ Torr base pressure. A 5 nm Se capping was deposited at room temperature to protect the film surface from oxidation under ambient conditions. The sample structure was monitored by XRD and STEM before and after recapping and showed no change in film quality. Due to partial removal of the Te capping before deposition of the new Se capping, the capping layer is assumed to be an oxidized Te-Se alloy, of which a diffraction peak~\cite{liu2024} is recorded in the reported diffractograms (Fig. \ref{fig:xrd}). We emphasize that the pristine film (also referred to as the as-prepared film) was likewise recapped. \add{The final stack is 10~nm \bt\ buffer, 60~nm \mbt\ film, and 5~nm capping.}

The recapped samples were irradiated using a Danfysik series 1090 high current implanter with a doubly charged argon beam (Ar$^{2+}$) at a 150 kV acceleration voltage (under an angle of 7 degrees off normal). Sample irradiation was modeled using SRIM-2013~\cite{ziegler2010srim} with monolayer collision steps and standard displacement energies for the lattice and surface. The ion and recoil distributions are reported in the Supplemental Material~\cite{SM}.

\subsection*{TEM sample preparation and STEM characterization}

Cross-sectional lamellae were extracted using a Focused Ion Beam (FIB) milling process (DualBeam FEI Nova 600 NanoLab). To preserve the integrity of the surface and sub-surface features, a dual-layer protective coating of electron-beam and ion-beam induced platinum was deposited prior to milling. Structural and chemical analysis were subsequently performed using an aberration-corrected STEM (cold-FEG JEOL ARM200F) operated at 200 kV. The system was equipped with a STEM-Cs corrector to achieve sub-angstrom resolution and a Centurio EDX spectrometer for high-sensitivity elemental mapping.

\subsection*{X-Ray Diffraction experiments and simulation}

Out-of-plane X-ray diffraction measurements were performed using a PANalytical X'Pert Pro diffractometer in Bragg geometry, employing Cu $K_\alpha$ radiation ($\lambda = 1.5406$ \AA{}). A Ni filter was used to suppress Cu $K\beta$ contributions. Diffraction patterns were collected using an X'Celerator detector over a fixed $2\theta$ range several times and averaged per sample. Samples were mounted on an amorphous glass holder that produces a slowly varying but consistent background in the diffractograms. Reference diffraction angles were extracted from structural parameters obtained from the Crystallography Open Database, entry 7210230. Simulated diffraction intensities were generated using the \texttt{pymatgen} Python module~\cite{ong2013}. Reported intensities account for the atomic scattering and Lorentz polarization factors. The intermixed structure referred to in the main text retains the same structure as the pristine material but has a homogeneous occupancy of all elements over all sites (while retaining the correct stoichiometry).

\subsection*{Computational Methods}

Density functional theory calculations were performed using the projector augmented wave (PAW) method~\cite{Kresse1996}, as implemented in the Vienna \textit{Ab initio} Simulation Package (VASP)~\cite{Kresse1999Jan}. The Perdew-Burke-Ernzerhof parameterization~\cite{Perdew1996Oct} of the generalized gradient approximation (GGA-PBE) was used to describe the exchange–correlation interaction. To characterize the localized Mn $3d$ states in \MBT, the DFT+$U$ method was applied with a Hubbard $U$ value of 4~eV, as implemented in the Liechtenstein scheme~\cite{Liechtenstein1995Aug}. The $U$ value was selected based on its alignment with previous first-principles studies~\cite{li2019,Li2020}. Long-range van der Waals interactions, essential for describing the interlayer distance between \MBT\ septuple layers, were considered using the DFT-D3 dispersion correction~\cite{Grimme2010Apr}. A plane-wave basis set with a kinetic energy cutoff of 350~eV was used following comprehensive convergence tests. To determine the defect energetics, a $3\times3\times1$ supercell was used with a $\Gamma$-centered $4\times4\times1$ $k$-point mesh and spin-orbit coupling (SOC). The total energy was converged to $10^{-6}$~eV in the self-consistent calculations. Structural relaxations were performed until the residual forces on the atomic positions and stresses were below 0.01~eV/\AA~neglecting SOC. A vacuum layer of 20~\AA\ was set along the $c$-axis to prevent spurious periodic interactions between layers. To investigate the structural properties of large-scale disordered systems that are beyond the computational limits of conventional DFT calculations, we employed machine-learning interatomic potentials (MLIPs) through the Atomic Simulation Environment (ASE)~\cite{Larsen2017}. We assessed several MACE (Machine Learning Interatomic Potentials) models~\cite{batatia2023macehigherorderequivariant} and selected the \texttt{MACE-medium-MPA-0} model~\cite{Batatia2025}. This model, pre-trained on the Materials Project (MPtrj)~\cite{Horton2025} and Alexandria datasets~\cite{SCHMIDT2024101560}, provided a reliable description of the Mn–Bi–Te chemical environment. As the base MACE potentials overestimated the vdW gap distance, DFT-D3 corrections were considered using the simple-d3 package~\cite{Ehlert2024, Grimme2010Apr}. The accuracy of the MACE+D3 approach was validated by benchmarking the vdW gap distance and the formation energies of Bi–Te swap defects with our DFT results (see \cref{fig:si-vdwgapdistances,fig:si-formationenergies} in Supplemental Material~\cite{SM}). The validated MACE+D3 framework was applied to a large-scale supercell containing 1120 atoms with a vacuum layer of 20~\AA . This model was designed to simulate the structural transitions between pristine, intermixed, and fully swapped bilayer configurations. The large-scale model was relaxed to a force criterion of 0.01~eV/\AA, enabling the evaluation of interlayer atomic distances.

\subsection*{Electrical transport \& Magnetometry}
Electrical Hall effect and magnetoresistance measurements are carried out in a Quantum Design PPMS DynaCool up to 14 T and down to 2 K. The excitation current is maintained at 100 $\mu$A. Rectangular samples cleaved from the GaAs wafer are measured in a 5-wire Hall configuration. A linear slope at high magnetic field is subtracted from the Hall resistance to obtain $R_{\text{AHE}}$. The conductivity is calculated as
\begin{align*}
    \sigma_{xx} = \frac{\rho_{xx}}{\rho_{xx}^2+\rho_{xy}^2}
    &\text{ and }
    \sigma_{xy}^{A} = \frac{\rho_{\text{AHE}}}{\rho_{xx}^2+\rho_{\text{AHE}}^2}\\
    \text{ with }
    \rho_{xx} = \frac{W}{L}R_{xx}t
    &\text{ and }
    \rho_{\text{AHE}} = R_{\text{AHE}}t,
\end{align*}
\noindent
where $W$, $L$, and $t$ are the sample width, length and thickness. Film thickness is taken as $75$ nm and includes the \bt\ buffer and capping.

Vibrating sample magnetometry was performed in a Quantum Design MPMS3 for in- and out-of-plane measurements under a DC field. \add{Field dependence was measured at temperatures ranging from 2 to 30 K in steps of 1 K.} The diamagnetic contribution of the substrate was subtracted by recovering the diamagnetic susceptibility from $M(H)$ curves at 150 K (above the N\'eel temperature of MBT). \add{ZFC and FC curves were measured upon heating with zero-field conditions set by oscillating the superconducting magnet before cooling.} 

XAS and XMCD measurements were carried out at the DEIMOS beamline of the SOLEIL synchrotron~\cite{ohresser2014}. The Mn L-edges were measured under normal incidence at 2 K with an out-of-plane magnetic field. A linear background was fitted from the pre-edge and subtracted from the XAS data. Standard sum-rule analysis was used to extract the effective orbital and spin moments with $n_{3d}=5$~\cite{thole1992, carra1993, chen1995}. The orbital moment is found to be $-0.11(5) \mu_B$ across all samples at low field and is excluded from the reported effective moments. The XMCD data serves to highlight that the change and trend in magnetization after irradiation can be attributed to Mn, and not to paramagnetic defects created by irradiation. As reported elsewhere the moments calculated by sum-rule analysis are lower than the expected moment of Mn in MBT~\cite{bibby2025,sun2024}. Exact treatment of the low $m^{\text{eff}}$ values is beyond the scope of this work and will be reported elsewhere. Factors influencing the extracted effective moment include (but are not limited to), mixing of $L_{2,3}$ edges due to low spin orbit coupling of Mn~\cite{piamonteze2009}, non zero-spin dipole~\cite{piamonteze2009}, self-absorption of the fluorescent X-rays~\cite{eisebitt1993}, and a non-trivial background from the Te $M_{4, 5}$ edges~\cite{tcakaev2023}.

\subsection*{Author Contributions}
R.J. performed all experimental measurements (excluding STEM/EDS and transport), analyzed all experimental data, and wrote the manuscript. H.X. performed  STEM/EDS measurements and contributed to analysis and writing. A.B.P.F. is the lead author for the atomistic modelling part, performed DFT calculations, and contributed to writing. S.B. performed transport measurements. S.K.B. assisted with XAS/XMCD and XRD. J.Z. contributed to DFT calculations. Z.Z., A.S.L., M.S., and P.O. assisted with XAS/XMCD measurements. X.L. grew the thin film samples by MBE. B.A.A. performed and analyzed transport measurements, supervised MBE sample growth, co-supervised R.J., and contributed to writing. B.G.M. and L.F. contributed to transport measurements. M.v.B. and C.M. co-supervised R.J. F.M.L. supervised H.X. J.C.C. supervised A.B.P.F. and contributed to the DFT work. J.W.S. performed and analyzed STEM/EDS measurements and supervised H.X. L.M.C.P. conceived and supervised the project. All authors reviewed and approved the final manuscript.
\subsection*{Acknowledgments}
This work was funded by the FWO and FRS-FNRS under the Excellence of Science (EOS) program (40007563-CONNECT) and by the KU Leuven (grant No. C14/21/08).
J. W. S. acknowledges the FWO infrastructure projects AKUL13/19 and I000920N.
S. K. B., S. B., X. L. and B. A. A. acknowledge support from National Science Foundation grant NSF-DMR-2313441. 
A. B. P. F. and J. C. C. acknowledge financial support from the F\'ed\'eration Wallonie-Bruxelles through the ARC Grant “DREAMS” (No. 21/26-116), and from the F.R.S.-FNRS through the research project No. T.029.22F.
Computational resources have been provided by the supercomputing facilities of the Universit\'e catholique de Louvain (CISM/UCL) and the Consortium des \'Equipements de Calcul Intensif en F\'ed\'eration Wallonie Bruxelles (C\'ECI) funded by the Fond de la Recherche Scientifique de Belgique (F.R.S.-FNRS) under convention No. 2.5020.11 and by the Walloon Region. The present research also benefited from computational resources made available on Lucia, the Tier-1 supercomputer of the Walloon Region, infrastructure funded by the Walloon Region under the Grant Agreement No. 1910247.
Synchrotron experiments were performed on the DEIMOS beamline at SOLEIL Synchrotron, France (proposal number 20230108). We are grateful to the SOLEIL staff for smoothly running the facility.

R. J. acknowledges the contributions of Bart Caerts for operation of the implanter during ion irradiation, Koen van Stipthout for helpful discussions during the conception of the work, and Stijn Reniers for assistance with diffraction experiments.

\subsection*{Data Availability}
The data that support the findings of this study are available
from the corresponding author upon reasonable request.

\bibliography{biblio_bibtex_abbrev}


\clearpage
\onecolumngrid           

\setcounter{section}{0}
\setcounter{figure}{0}
\setcounter{table}{0}
\setcounter{equation}{0}
\setcounter{page}{1}
\renewcommand{\thesection}{\Roman{section}}
\renewcommand{\thefigure}{S\arabic{figure}}
\renewcommand{\theHfigure}{SI.\arabic{figure}}  
\renewcommand{\thetable}{S\Roman{table}}
\renewcommand{\theHtable}{SI.\arabic{table}}
\renewcommand{\theequation}{\arabic{equation}}
\renewcommand{\theHequation}{SI.\arabic{equation}}

\begin{center}
    {\Large\textbf{Supplemental Material}}\\[0.4em]
    {\large Disorder-driven symmetry suppression by van der Waals planar
    defects in a magnetic topological insulator}\\[0.8em]
    {\normalsize
    Rikkie Joris,$^{1,*}$
    Heyi Xia,$^{2}$
    Ana Beatriz Pedro Fontes,$^{3}$
    Sara Bey,$^{4}$
    Seul-Ki Bac,$^{5}$
    Jiaqi Zhou,$^{3}$
    Zviadi Zarkua,$^{1}$
    Ahmed Samir Lotfy,$^{1}$
    Muhammad Saad,$^{1}$
    Philippe Ohresser,$^{6}$
    Bence G. M\'{a}rkus,$^{4,7}$
    L\'{a}szl\'{o} Forr\'{o},$^{4,7}$
    Margriet van Bael,$^{1}$
    Clement Merckling,$^{2,8}$
    Xinyu Liu,$^{4}$
    Francisco Molina-Lopez,$^{2}$
    Jean-Christophe Charlier,$^{3}$
    Jin Won Seo,$^{2}$
    Badih A.\ Assaf,$^{4}$
    and Lino M.\ C.\ Pereira$^{1}$
    }\\[0.8em]
    {\small
    $^{1}$Quantum Solid State Physics, KU Leuven,
    Celestijnenlaan 200D, B-3001 Leuven, Belgium\\[0.2em]
    $^{2}$Department of Materials Engineering, KU Leuven,
    Kasteelpark Arenberg 44, B-3001 Leuven, Belgium\\[0.2em]
    $^{3}$Institute of Condensed Matter and Nanosciences,
    Universit\'{e} Catholique de Louvain,
    B-1348 Louvain-la-Neuve, Belgium\\[0.2em]
    $^{4}$Department of Physics, University of Notre Dame,
    Notre Dame, Indiana 46556, USA\\[0.2em]
    $^{5}$Institut des NanoSciences de Paris, INSP,
    Sorbonne Universit\'{e}, CNRS,
    4 place Jussieu, F-75005 Paris, France\\[0.2em]
    $^{6}$Synchrotron SOLEIL, L'Orme des Merisiers,
    F-91190 Saint Aubin, France\\[0.2em]
    $^{7}$Stavropoulos Center for Complex Quantum Matter,
    University of Notre Dame, Notre Dame, Indiana 46556, USA\\[0.2em]
    $^{8}$IMEC, B-3001 Leuven, Belgium\\[0.4em]
    $^{*}$Contact author: rikkie.joris@kuleuven.be
    }
\end{center}

\section{Ion Distribution and Displacements per Atom}\label{SI:beam}

\begin{figure}[h]
    \centering
    \includegraphics[width=\linewidth]{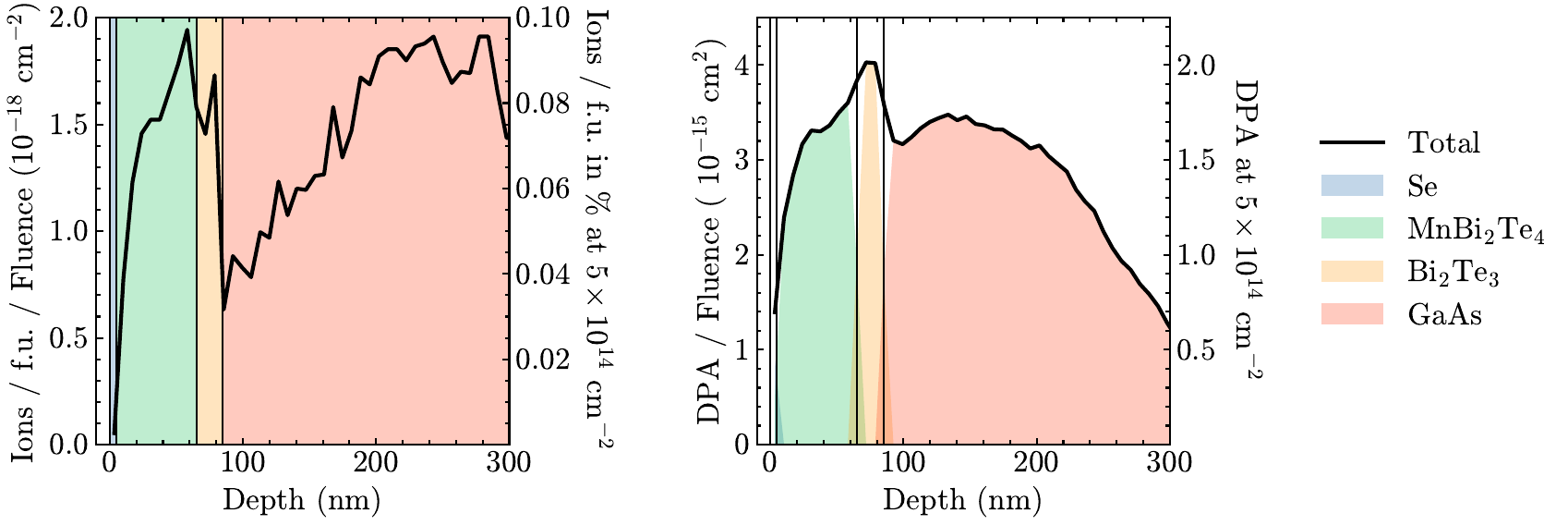}
    \caption{\textbf{Left}: Argon ion density per fluence as simulated by SRIM plotted in function of depth throughout the thin film stack. The Ar ion density is rescaled to a concentration (ions per formula unit) for each material in the stack. For practical reference the right hand side vertical axis denotes the Ar ions per formula unit for a fluence of \fluence{5}{14} (the highest fluence reported in this work). The maximal concentration within the \mbt\ layer at this fluence is less than $0.1 \%$ Ar per unit cell. \textbf{Right}: The displacements per atom (DPA) within the thin film stack as simulated by SRIM per fluence. The DPA is calculated by dividing the total recoil density by the number density (of all atoms) per material in the stack. For reference the DPA for a fluence of \fluence{5}{14} is given on the right vertical axis.}
    \label{fig:si-dpa}
\end{figure}

During ion irradiation, the projectile atoms lose energy during inelastic scattering events. Atoms in the crystal lattice are displaced from their lattice sites by the projectile atoms through collisions and on their turn scatter until their kinetic energy is sufficiently reduced to either (re)combine with a vacancy or bind as an interstitial atom in the lattice structure. As such, defects created by ion irradiation are inherently zero dimensional point defects (antisites, vacancies, and interstitials). Within the thin film, the energy required to displace an atom from its lattice site is many orders of magnitude lower than the kinetic energy of the projectile atoms (here, 20-30 eV compared to 300 keV), i.e. all atoms are displaced at roughly an equal rate. The creation and final population of defects is governed by their formation energy and their dynamics during the (non-equilibrium) process of irradiation, not by preferential scattering.

We emphasize that this work pertains to effects of ion \textit{irradiation} not ion \textit{implantation}. Implantation occurs when the projectile atoms lose all kinetic energy within the thin film and themselves bind as a point defect in the crystal lattice. Irradiation occurs when the majority of the projectile atoms are implanted in the substrate, while external doping by the projectile atoms in the film is negligible. As shown in Fig. \ref{fig:si-dpa} we estimate that an Ar beam of 300 keV at a fluence of \fluence{5}{14}\ dopes the MBT film to a level of 0.001 Ar atoms per unit cell. The doping of Ar into the film at the low fluence discussed in the main text lies one order of magnitude below that. 

Atoms displaced from their lattice sites are referred to as recoils. When recoils from one film in the stack traverse the interface and are implanted in a neighboring film in the stack, this is referred to as kinetic intermixing. As is shown in Fig. \ref{fig:si-dpa}, we estimate kinetic intermixing of the \mbt\ film and \bt\ buffer to be limited to a distance of $\sim 10\ \text{nm}$ orthogonal to the interface. Inspection of the \mbt\ structure in the middle of the film  therefore allows us to ignore effects of kinetic intermixing. The cation intermixing and novel disordered structure discussed in the main text can be fully attributed to scattering events within the \mbt\ film. Due to the limited thickness of the \bt\ film, kinetic intermixing is however expected to play a major role in the modification of the buffer layer.

The metric used to compare ion beam damage across materials is the displacements per atom (DPA). The local density of recoils is scaled to the local number density of the material to recover the (average) amount of times atoms were displaced at a given fluence. The recoil distribution (the amount of displacements) varies throughout the film as projectile atoms lose energy while they traverse the stack. As is evident from Fig. \ref{fig:si-dpa} the DPA near the surface of the \mbt\ film is lower than the DPA in the middle of the film. This surface region is consequently excluded from reported results. We emphasize that the DPA can be interpreted literally. At the highest fluence reported in this work, a periodic order is still found, while we estimate that the atoms within a single \mbt\ unit will each have been displaced from their lattice site $\sim 1.6$ times (on average).    

\section{Large scale micrographs}

\begin{figure}[h]
    \centering
    \includegraphics[width=7in]{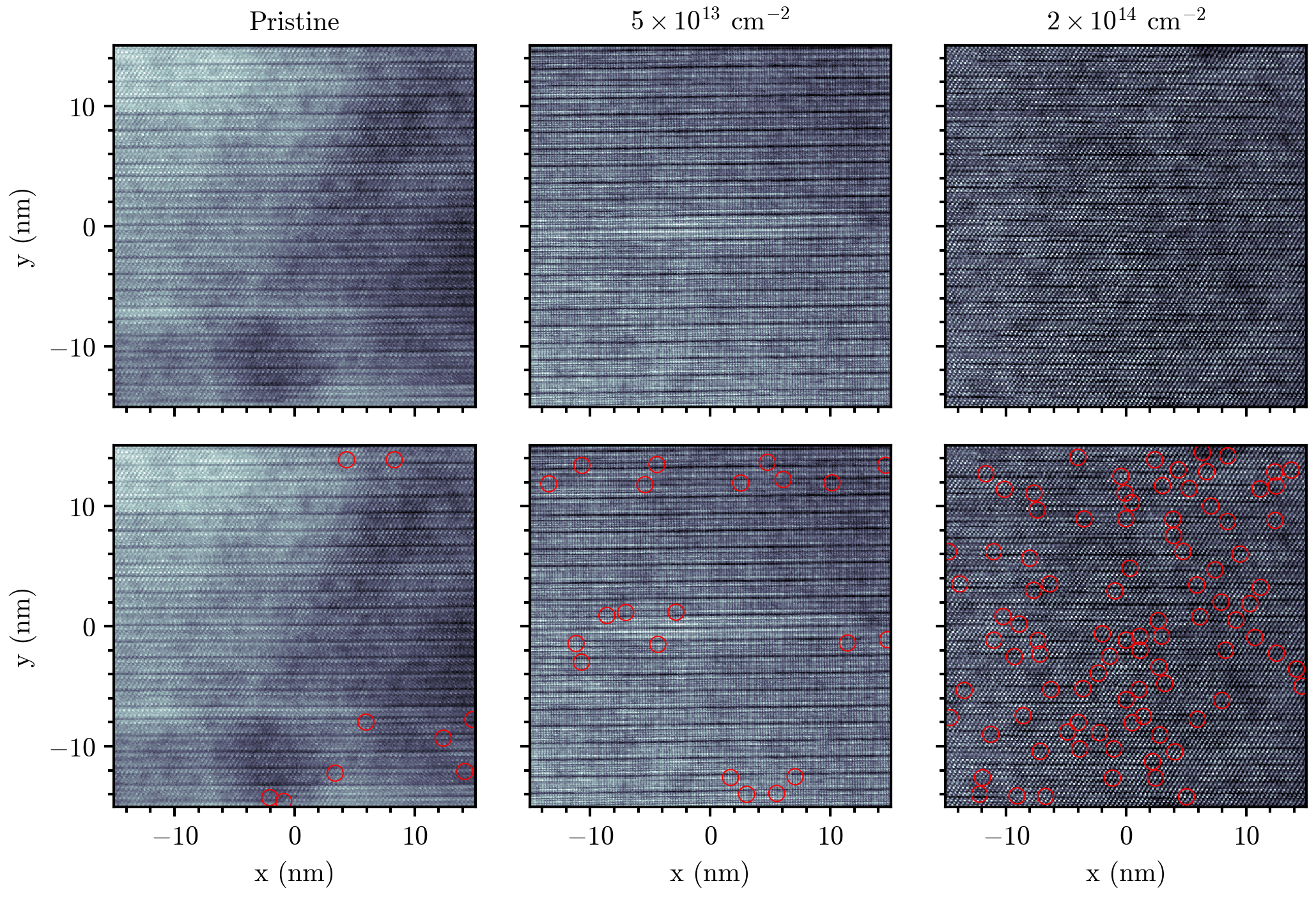}
    \caption{\add{Micrographs of the pristine, low fluence, and high fluence irradiated films (left to right) obtained in the middle of the films. Annotated on the bottom are points where a bilayer detaches or attaches to a vdW layer as identified by inspection.}}
    \label{si:fig:count}
\end{figure}

\add{Figure \ref{si:fig:count} contains three $30\times30$~nm micrographs taken in the middle of the pristine and irradiated films. As discussed in the main text, for the as-prepared and low fluence irradiated film, the vdW gap is well defined. After irradiation by the high fluence, this is no longer the case, due to the high density of planar defects. We draw two additional conclusions from these large scale micrographs.}

\add{Firstly, the intercalated \bt\ layers survive irradiation by the low fluence. As seen in the center of the micrograph, two QLs are present between the \mbt\ vdW layers. We therefore see no direct indication that ion irradiation at this fluence dissolves the intercalated \bt\ phase into the \mbt\ film through kinetic intermixing. The intensity of the \bt\ diffraction peaks reported in Fig. \ref{fig:xrd} can therefore not be attributed to either the buffer layer or intercalated layers alone up to (and including) a fluence of \fluence{5}{13}.}

\add{Secondly, the micrographs show the observed planar defects are ubiquitously present throughout the film, and are not a structural distortion that can only be observed locally. We have annotated points in Fig. \ref{si:fig:count} where a bilayer detaches from or joins a vdW layer. The presence of several planar defects in the as-prepared film does not in itself show the defects form spontaneously after growth. Preparation of the lamellae used for STEM may induce cation-anion intermixing, which result in these planar defects. The observed defect density therefore sets an upper bound on the number of planar defects that can be generated by the preparation procedure. The high concentration of planar defects in the high fluence irradiated film can therefore not be attributed to this procedure.}

\add{To indicate the defect density, the number of vdW gaps can be counted that run uninterrupted by a planar defect from left to right over the domain of the micrograph (30~nm). In the as-prepared micrograph this constitutes 17 gaps, in the low fluence 13 gaps, and 0 gaps in the micrograph after high fluence irradiation. The number of points where a bilayer attaches to or detaches from a vdW layer can be normalized to the micrograph area, giving 0.01, 0.02, and 0.1~nm$^{-2}$ for the as-prepared, low fluence, and high fluence irradiated micrographs respectively. The reported values must strictly be interpreted as indicative, as the definition of detach and attach is ambiguous. Nonetheless, the increase of planar defects observed in the low fluence irradiated film as compared to the as-prepared film indicates some cation-anion intermixing and defect self-organization is already present at that fluence. This implies that the density of planar defects may be tunable by fluence between the low and high fluences.}

\section{Directionality of dislocations and symmetry suppression}\label{si-disloc}

\begin{figure}[h]
    \centering
    \includegraphics[width=0.75\linewidth]{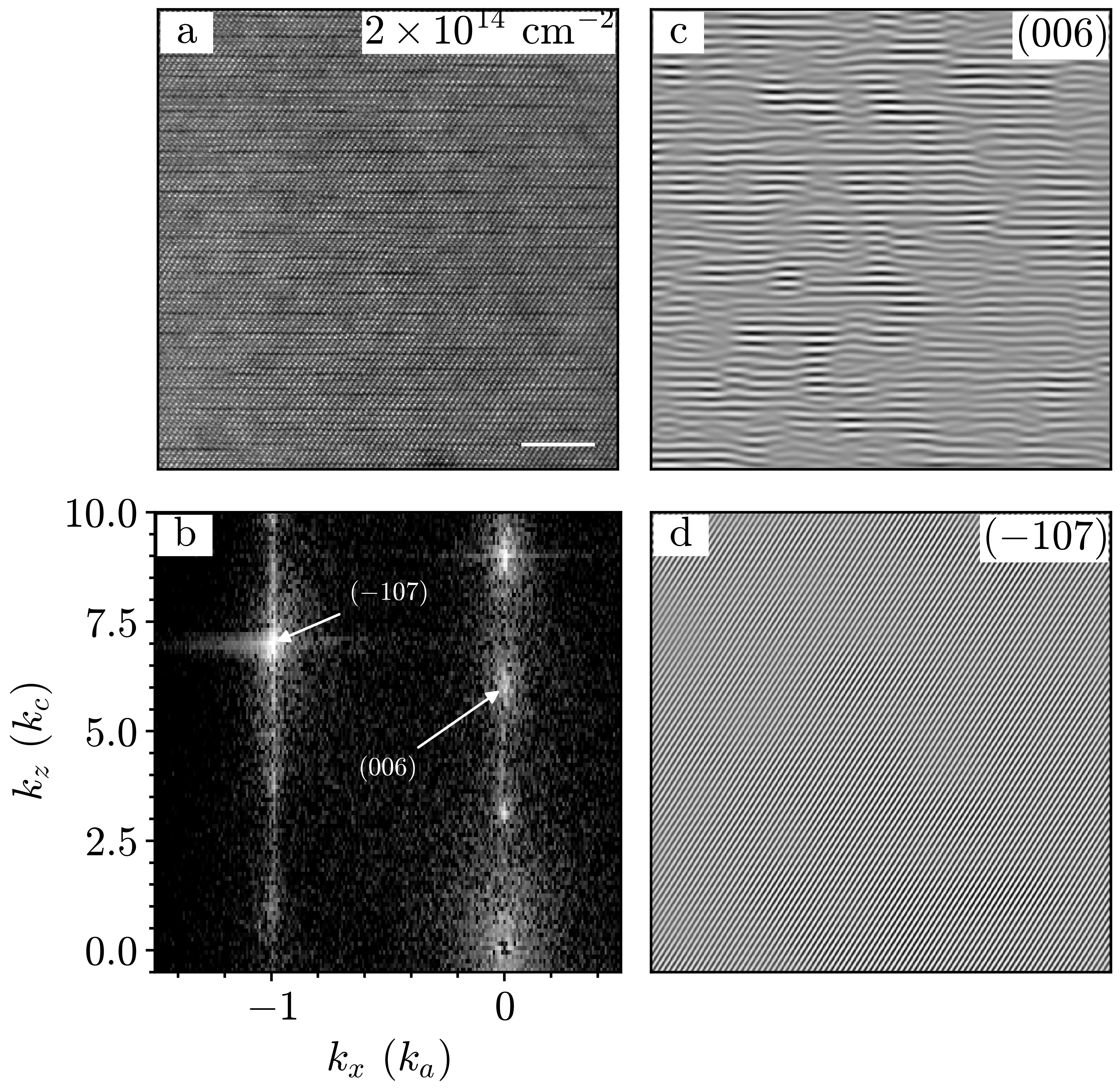}
    \caption{(a) Large scale STEM micrograph obtained in the middle of the MBT thin film. Scale bar represents 5 nm. (b) Logarithm of the modulus squared of the discrete Fourier transform generated from the micrograph (a). The reciprocal space is expressed as multiples of the reciprocal lattice constant in the relevant direction. (c, d) Real component of the inverse discrete Fourier transform containing only contributions from the $(006)$ and $(-107)$ peak respectively. Note that due to the small spacing of intensity variations in (d), it is advisable to view the figure enlarged digitally to avoid effects of image compression and printing.}
    \label{fig:SI_largeDFT}
\end{figure}

The formation of planar defects due to self-organization of cation-anion antisite defects results in a remarkable feature: the planar disorder only generates dislocations in the $ab$-plane and leaves other crystallographic directions devoid of dislocations. Figure \ref{fig:SI_largeDFT} highlights a large scale micrograph of the MBT film after high fluence irradiation. As discussed in the main text, the dominating form of structural disorder are the vdW specific planar defects (swapped bilayers and ripplocations). The existence of clear peaks along the crystallographic directions in the discrete Fourier transform of the micrograph (Fig. \ref{fig:SI_largeDFT}) reinforces the interpretation that the ion-beam induced disorder changes the periodic order of the crystal structure but does not disrupt it entirely. 

As annotated in the figure, two crystallographic directions are of interest. Firstly, the $(006)$-direction which corresponds to half a SL separation along the $c$-axis. The intensity of the Fourier transform at this length scale is sensitive to both occupation of the vdW gap by planar defects and adaptation of the layer separation. The inverse Fourier transform of this peak (Fig. \ref{fig:SI_largeDFT}(c)) shows a high concentration of dislocations that can be correlated to presence of planar defects in the micrograph. Secondly, in stark contrast, the inverse Fourier transform of the $(107)$ peak contains zero dislocations (Fig. \ref{fig:SI_largeDFT}). This crystallographic direction aligns with the diagonal planes that contains all seven atoms within a SL. Remarkably, even when a detached bilayer is present between two vdW layers, the atoms within the detached bilayer still align sufficiently with the diagonal planes to not form dislocations in the crystallographic direction.

This behavior can be understood from the microscopic model presented in the main text. The formation of the planar defects due to the self-assembly of point defects shifts the detaching bilayer away from its parent vdW layer, but does not buckle it. The planar defect therefore presents as a dislocation in the $ab$-plane, strongly interfering with periodic order along the $c$-axis (displacement from parent vdW layer), but leaves the $(107)$-direction largely unchanged (absence of buckling). The disordered phase generated by high fluence irradiation therefore selectively suppresses crystallographic symmetry along the $c$-axis, while other symmetries within the crystal are retained.

\section{Additional Structural Characterization}\label{SI:structure}

\begin{figure}[h]
    \centering
    \includegraphics[width=\linewidth]{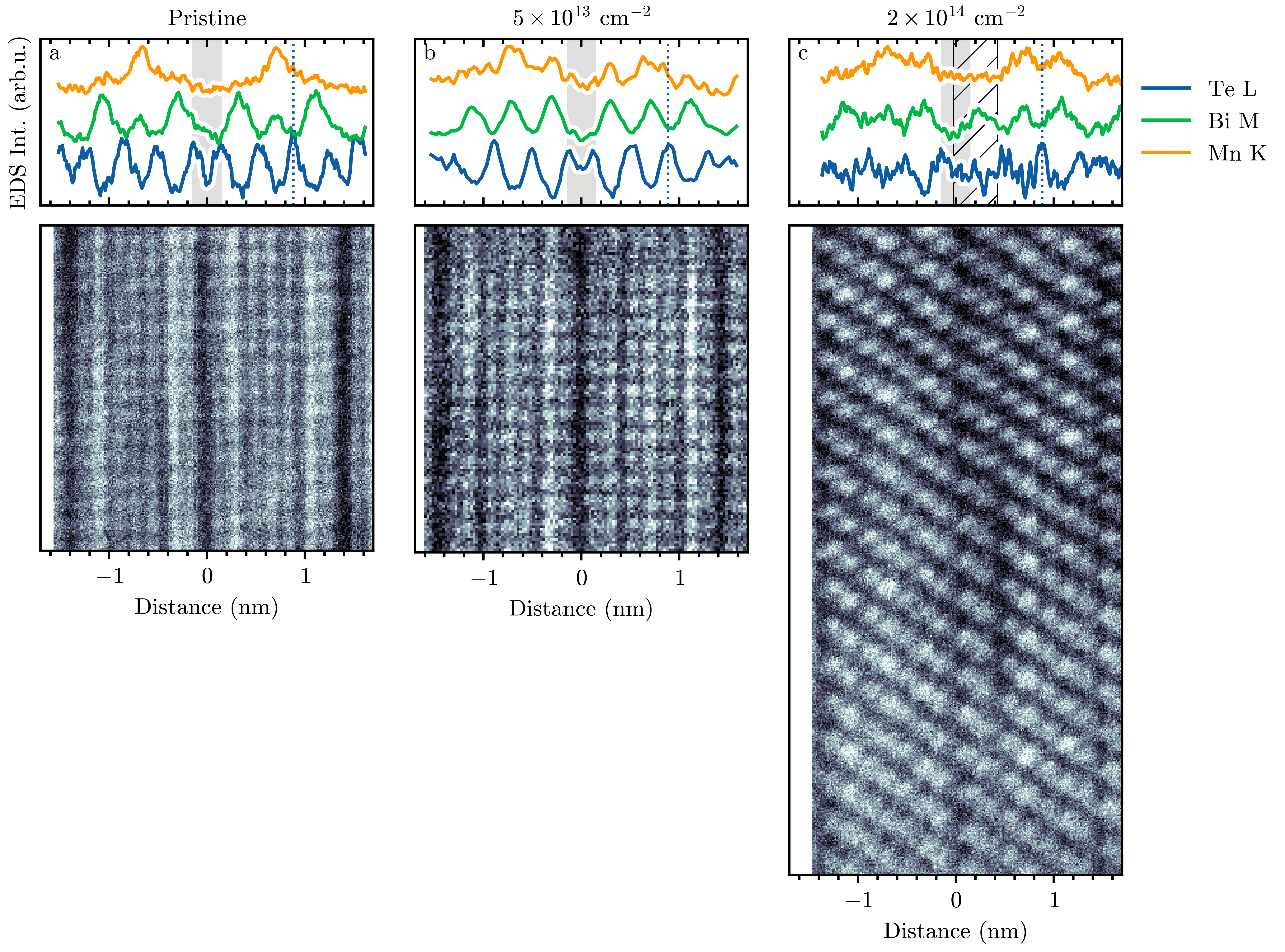}
    \caption{Complete set of EDS data partially shown in Figure \ref{fig:tem-edx} for all fluences as indicated in the figure titles. The dotted blue line indicates the Te layer to which the signals were aligned spatially. The shaded gray area is the region between the terminating Te layers (spanning the vdW gap) in the pristine film. Hatched area covers the detached bilayer. \add{Micrograph reported in Fig. \ref{fig:tem-edx} are reproduced here.}}
    \label{fig:si-edxlines}
\end{figure}

As discussed in the main text, clear observations and conclusions can be made from the cation EDS signals, as reported. Features in the Te L-edge signal of the high fluence irradiated film are less easily discerned. The implications of the Te related EDS results are thus presented here. Figure \ref{fig:si-edxlines} shows the spatially integrated EDS intensities corresponding to the micrographs given in Fig. \ref{fig:tem-edx}. Both pristine (Fig. \ref{fig:si-edxlines}(a)) and low fluence irradiated samples (Fig. \ref{fig:si-edxlines}(b)) contain a clear separation of lattice planes, while the high fluence irradiated sample (Fig. \ref{fig:si-edxlines}(c)) contains greater chemical disorder.   

\subsection{Chemical Nature of the Ripplocation}

As discussed in the main text, the ripplocation is found to contain Bi with little to no Mn. As shown in Fig. \ref{fig:si-edxlines}(c) the spatially integrated EDS curves (partially reported in Fig. \ref{fig:tem-edx} in the main text) reveal that the Mn EDS intensity drops to near background levels in the vdW gap (within the noise). Likewise, there is no correlation between the Mn and Bi EDS curves in this region. The Bi and Te EDS curves do not show this behavior. The presence of Bi in the gap and ripplocation is clear. The Te signal contains several minor peaks in the gap (grey area) and ripplocation region (hatched area). Due to the noise contained in the Te EDS curve, the peaks can not be clearly discerned and correlated to the Bi curve. It is however evident that the Te EDS curve does not decrease to background levels within this region. We conclude that the detached bilayer consists of Bi and Te atoms with a negligible Mn concentration. 

\subsection{Additional Defects: \texorpdfstring{\fluence{5}{13}}{5E13 cm-2}}

Cation antisites are found to be the dominant form of disorder in the low fluence irradiated film while the long-range crystallographic order of the system is maintained. Figure \ref{fig:si-edxlines}(b) shows that beside cation antisites, cation-anion intermixing and Te interstitials are also present. The Te EDS signal does not decrease to background levels inside the vdW gap (region indicated by shaded area). The presence of Te atoms within the vdW gap likewise implies the presence of vacancies in the SL layers. The Bi and Mn EDS curves show a presence of \mnte\ and \bite\ antisite defects in the terminating Te layers of the vdW layers. Both cation signals however decrease to background levels in the vdW gap. The presence of Te interstitials in the vdW gap shows the SL have a stoichiometry slightly different to the pristine layers. We emphasize that, as the Te population is still present locally and not displaced from the system, the (middle of the) film should be regarded as compositionally identical to the pristine film. The magnitude of Te vacancy and interstitials can not be estimated. Vacancies left behind by the Te found in the vdW gap can be filled by either Mn, Bi, or Te atoms. An absence of Mn and Bi in the vdW gap is noted, and cation-anion antisites are found: the vacancy is likely not limited to Te sites and propagates to both Mn and Bi sites. Importantly, the corresponding micrograph (Fig. \ref{fig:tem-edx}(g)) does not contain atomic columns in the vdW gap. The interstitial Te atoms are likely bound to the surface of the SL layers over multiple different sites~\cite{netsou2020,wang2011}.  

As highlighted in the main text, the recorded vacancy and cation-anion intermixing in the low fluence irradiated film are present to a lesser concentration than defects related to cation site disorder. The presence of a broad spectrum of point defects does imply a complex dependence of transport properties on the fluence, as the recorded defects contain both donors and acceptors~\cite{du2021}. Irradiation at lower fluences is therefore not guaranteed to monotonically modify the majority carriers from $p$- to $n$-type.

\subsection{Additional Defects: \texorpdfstring{\fluence{2}{14}}{2E14 cm-2}}

\begin{figure}[h]
    \centering
    \includegraphics[width=\linewidth]{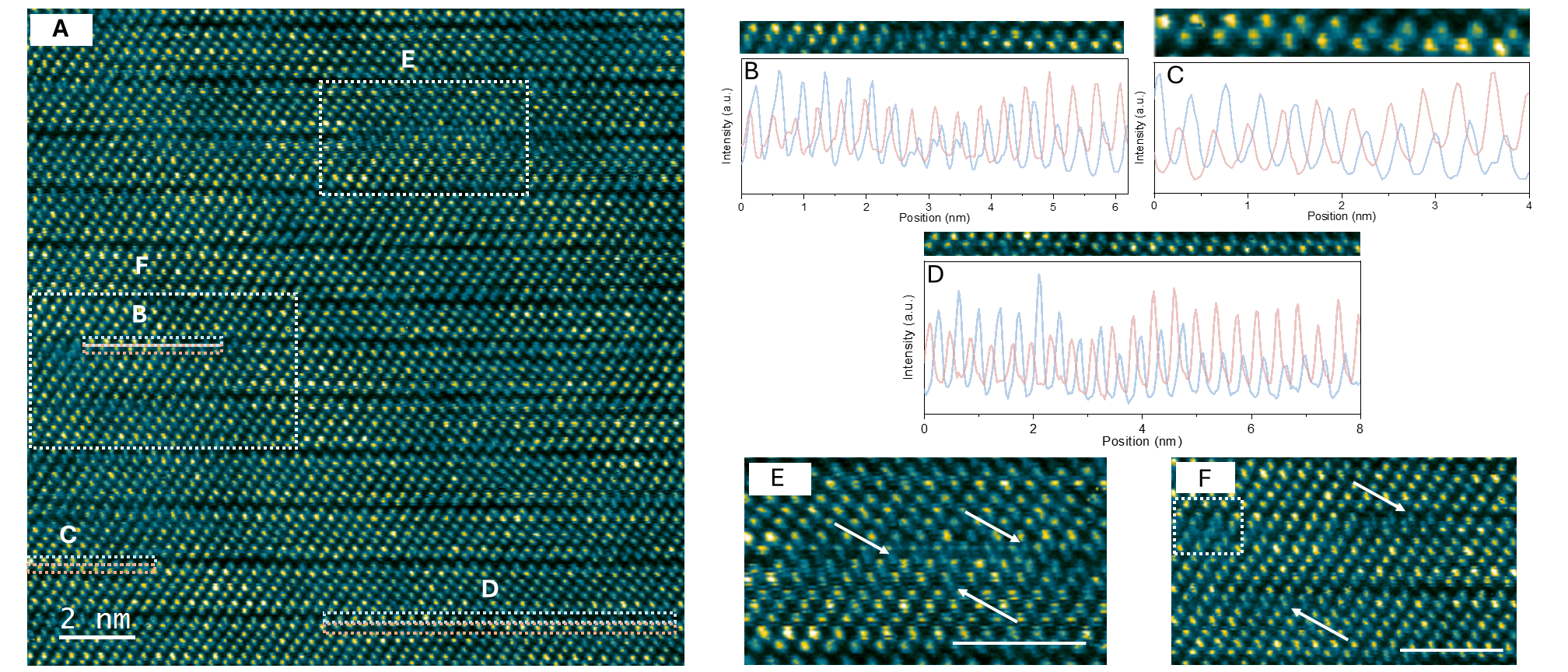}
    \caption{STEM micrographs showing swapped bilayers and interstitials.}
    \label{fig:si-interstitials}
\end{figure}

After irradiation with a high fluence the periodic SL structure is modified to a highly disordered one. We show these defects form due to self organization of cation-anion antisite defects. These planar defects are not the only defects found in the structure, but are the vast majority of the recorded defects (as further discussed below). Figure \ref{fig:si-interstitials}(a) contains a micrograph with several examples of defects, they are found to be either related to interstitial atoms or are variations on the swapped bilayer/ripplocation defect. Several swapped-bilayer-type defects are shown in Fig. \ref{fig:si-interstitials}(b-d) with their corresponding STEM intensity per layer plotted beside them. The lateral range over which the bilayer swaps is linked to the abruptness of the intensity change between the top and bottom layer of the bilayer. As visible in the swapped bilayer in Fig. \ref{fig:si-interstitials}(b) (and also indicated on Fig. \ref{fig:si-interstitials}(f)) additional atomic columns can be seen between the detached bilayer and the neighboring quintuple layer. Figure \ref{fig:si-interstitials}(e) highlights two SLs between which the vdW gap is occupied by a ripplocation and interstitial atomic columns. The combined structure of planar and interstitial defects fully bridge the vdW gap. We note three sites where interstitials occur, one in the terminating Te (anion) plane of the top SL, another adjacent to this terminating anion plane, and finally between (nomenclature from the pristine structure) the Te$_2$ and Bi plane of the bottom vdW layer. The observation of interstitials in these sites is in line with previous reports in the \ce{(V)_2(VI)_3} family~\cite{callaert2020} or have been predicted to exist as stable sites for 3d TM dopants in \st\ ~\cite{wang2024}.

This indicates the formation of stable interstitials that coexist with the detached bilayers. The combination of interstitials and swapped bilayers has not been reported in literature, their existence is however readily understood from the mechanisms presented in the main text. After (or during) ion bombardment the cation-anion antisites agglomerate to form higher dimensional defects. These complex microstructures thereafter offer new sites to which interstitial atoms can bind or migrate to. As swapped bilayers and ripplocations are recorded without the presence of interstitials, and when they are present the whole range of the planar defect is not always covered by interstitials; we conclude that these interstitials are a parallel pathway for the structure to alleviate cation-anion intermixing, and not a necessary factor for the formation of the planar defects. Interstitial defects in the \ce{(V)_2(VI)_3} family are known to be hyper mobile, and have been recorded crossing the vdW gap in real time~\cite{callaert2019}. We consider it likely that the swapped bilayer and ripplocation defects operate as accumulation sites for these mobile interstitials.

Gap-filling interstitials are observed with minor frequency. We emphasize that on large scales, as depicted in Fig. \ref{fig:SI_largeDFT}, the swapped bilayer and ripplocation defects are the dominant form of disorder, and occur at high concentrations. Likewise, the $(107)$-direction remains devoid of dislocations and no amorphous regions are observed.

\clearpage

\section{Validation and Benchmarking of Machine Learning Potentials}\label{si-computationaldetails}

\begin{figure}[ht]
    \centering
    \includegraphics[width=0.6\linewidth]{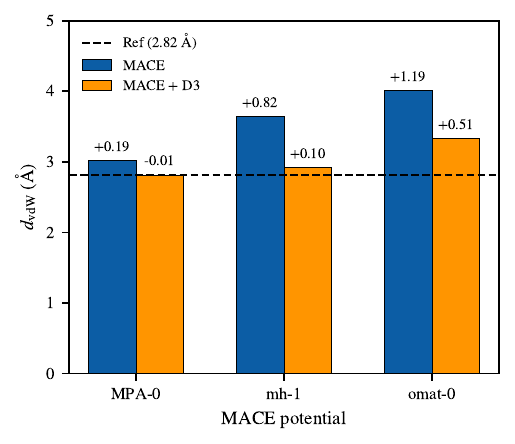}
    \caption{Comparison of the calculated vdW gap distance ($d_{\text{vdW}}$) for the $\mathrm{MPA}$-$0$, $\mathrm{mh}$-$1$, and $\mathrm{omat}$-$0$
    models without (blue) and with DFT-D3 corrections (orange). The horizontal dashed line represents the DFT reference value of 2.82 \AA. The numerical labels above the bars indicate the deviation from the reference value.
}
    \label{fig:si-vdwgapdistances}
\end{figure}

\begin{figure}[ht]
    \centering
    \includegraphics[width=\linewidth]{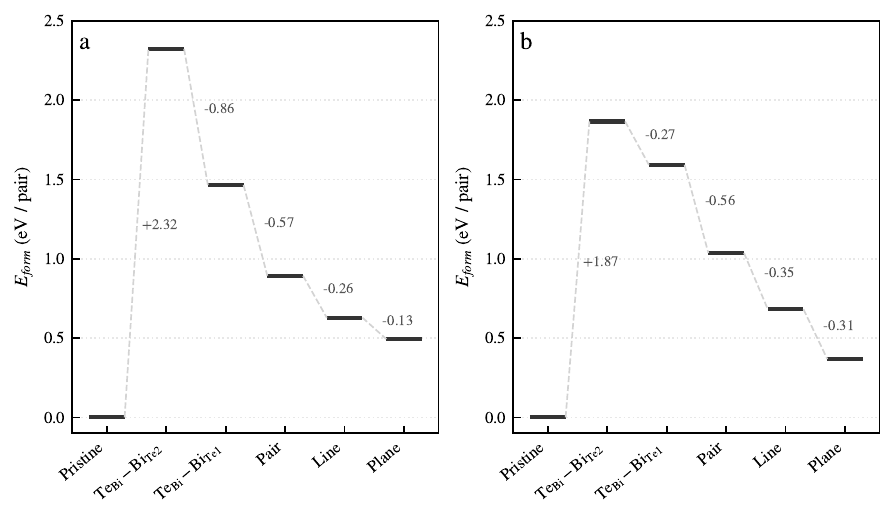}
    \caption{Comparison of defect formation energy landscapes. Calculated formation energy ($E_{\text{form}}$) in units of eV per pair for each structural configuration using (a) DFT and (b) the MPA-0+D3 machine-learning potential. The pristine system is defined as the energy reference (0 eV). Dashed gray lines and numerical labels indicate the energy differences ($\Delta E$) between successive defect states, illustrating the energetic stabilization as defects aggregate from isolated antisite positions into ordered line and planar structures.
}
    \label{fig:si-formationenergies}
\end{figure}

To validate the machine learning interatomic potential (MLIP) for the \MBT system, we compared it against first-principles DFT calculations. \Cref{fig:si-vdwgapdistances} shows that although the pre-trained MACE potentials overestimated the interlayer van der Waals spacing, incorporating the DFT-D3 corrections significantly improves the accuracy. Specifically, the MPA-0+D3 model reproduces the DFT reference vdW gap ($2.82$~\AA) with a deviation of less than $0.01$~\AA. \Cref{fig:si-formationenergies} compares the defect formation energy landscapes of Bi–Te swapped antisite defects. The MACE-MPA-0+D3 potential accurately captures the qualitative energetic trends and the stabilization observed as defects aggregate from isolated positions into ordered structures. 
The consistency in $\Delta E$ and the accurate reproduction of the vdW gap confirm that the potential provides a reliable potential energy surface for structural optimization of the MBT system.

\newpage

To quantify the variation in interlayer distance along the $x$-direction of the relaxed supercell, we employed a localized averaging scheme. Firstly, we sorted the atoms according to their vertical $z$-coordinate. The atoms were grouped into discrete atomic layers, denoted as $L_n$, by identifying sequential atoms whose $z$-coordinates fell within a tolerance $\tau = 0.4$~\AA:
\begin{equation}
    |z_{k} - z_{k-1}| \leq \tau
\end{equation}
where $z_k$ is the vertical position of the $k$-th atom in the initially sorted list. 
Secondly, the lattice vector $\mathbf{a}$ was discretized into $N = 17$ uniform bins. For each bin $j \in \{1, \dots, N\}$, the boundaries in fractional coordinates $(x'_j, x'_{j+1})$ are defined as:
\begin{equation}
    x'_j = \frac{j-1}{N}, \quad x'_{j+1} = \frac{j}{N}
\end{equation}
The corresponding real-space position along the $x$-axis for each bin, $X_j$, is taken at the bin center:
\begin{equation}
    X_j = \frac{a}{2}(x'_j + x'_{j+1})
\end{equation}

For a given pair of adjacent layers $L_n$ and $L_{n+1}$, the local interlayer spacing $d_{n, n+1}(X_j)$ within the bin $j$ was calculated as the difference between the mean $z$-coordinates of the atoms belonging to that bin. Let $S_{n,j}$ be the set of $z$-coordinates of atoms in layer $L_n$ whose fractional $x$-coordinates $x'_k$ fall within the boundaries of bin $j$:
\begin{equation}
    S_{n,j} = \{ z_k \in L_n \mid x'_j \leq x'_k < x'_{j+1} \}
\end{equation}
The local average height $\bar{z}_{n,j}$ for layer $n$ in bin $j$ is calculated as:
\begin{equation}
    \bar{z}_{n,j} = \frac{1}{|S_{n,j}|} \sum_{z \in S_{n,j}} z
\end{equation}
The local interlayer distance $d_{n, n+1}$ at position $X_j$ is then defined as:
\begin{equation}
    d_{n, n+1}(X_j) = \bar{z}_{n+1,j} - \bar{z}_{n,j}
\end{equation}
This procedure was applied specifically to the gaps between layers 7--8, 8--9, and 9--10 to capture the deformation profile between the MBT atomic layers.

\section{Field-dependent magnetization at various temperatures}
\begin{figure}[ht]
    \centering
    \includegraphics[width=\linewidth]{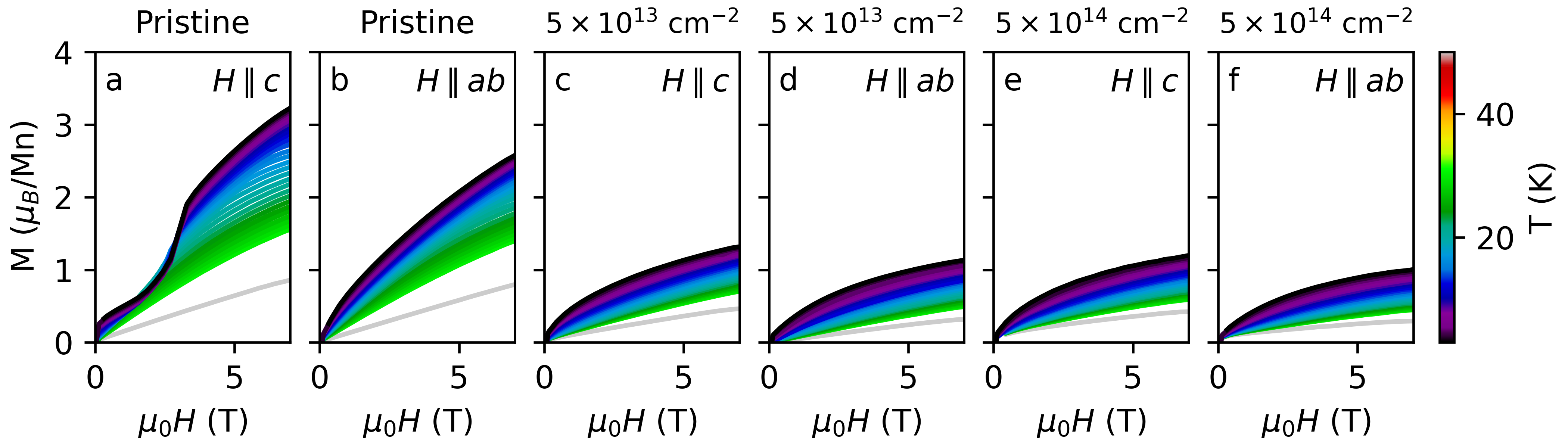}
    \caption{\add{(a--f) $M(H)$ curves used to calculate $M^{c}-M^{ab}$ in Fig. \ref{fig:MH} and $\chi_v=\text{d}M/\text{d}H$ in Fig. \ref{fig:MT} in the main text. Fluence is indicated on top, and annotation in top-right indicates applied field direction.}}
\end{figure}

\section{Additional Magnetometry data and interpretation}\label{si:dmdt}

\add{This section further discusses the $M(T)$-curves reported in Fig. \ref{fig:MT}. The temperature dependence of the magnetization alone is not able to uniquely identify the exact magnetic ground state in a system. Ferromagnetic, antiferromagnetic, glassy, and cluster-like transitions can produce similar $M(T)$-curves. Within the context of this manuscript, the exact magnetic ground state of the films after ion irradiation is not of interest. As given in the main text: we provide a basic description of the macroscopic magnetic properties (such as the reduction of the as-measured macroscopic magnetic anisotropy and the absence of field-driven transitions) after ion irradiation.}

\add{Additional information can be extracted about the magnetic order that is not available from $M(H)$ data alone. A difference between ZFC and FC curves (bifurcation) across a magnetic transition occurs when certain spin configurations can be made metastable under the presence of an applied field. In other words, the applied field strength and temperature alone do not uniquely determine the magnetic state the system is in: there are irreversible processes that depend on the presence of a magnetic field when cooling over the transition temperature. There are energetic barriers that separate spin configurations, if the magnetic field is able to overcome those (Zeeman energy), no irreversibility and bifurcation is seen. The magnitude of the bifurcation, at what field strength and in what direction, therefore holds information on these energetic barriers, and the microscopic parameters that determine the magnetic ground state, as is further argued in the main text.}

\begin{figure}[ht]
    \centering
    \includegraphics[width=\linewidth]{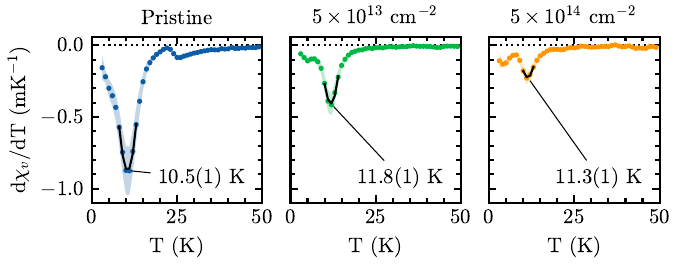}
    \caption{\add{The (numerical) temperature derivative of the susceptibility $\chi_v$ per fluence. The susceptibility measured with a 50 mT out-of-plane field under field-cooled conditions (Fig. \ref{fig:MT}a) is binned and differentiated with respect to temperature. Shaded regions indicate error estimates of the derivative based on the standard deviation of $\chi_v$ per bin. A region around the minimum of each d$\chi_v/$dT curve is fitted with a parabola to determine the transition temperature. Errors on the transition temperature are taken as the (original) measured temperature step. The N\'eel transition is visible in the pristine panel.}}
    \label{si:fig:dmdt}
\end{figure}

\add{In the as-prepared film, this low temperature transition can be linked to \mnbi\ antisite spin freezing \cite{sahoo2024} (see especially Fig. S7 therein). Below $T_{\text{N}}$ but above the freezing transition $T^*$, the \mnmn\ spins form an interlayer AFM order, while the \mnbi\ antisite spins remain disordered. Only below $T^*$ do these spins freeze antiparallel to the \mnmn\ spins in their vdW layer. This freezing was discovered by muon spin spectroscopy, but is shown to also provide a distinct feature in $M(T)$-curves at 11.5(1) K in powders. Figure \ref{si:fig:dmdt} shows the low temperature transition in the pristine film occurs at $10.5(1)$ K. After ion irradiation, the transition temperature moves, with $11.8(1)$ K after low fluence irradiation, and $11.3(1)$ K in the high fluence regime. As discussed in the main text, the N\'eel transition is no longer observed in the irradiated samples.}

\add{It is interesting that the low temperature transition, occurs at similar temperature throughout the irradiation series. This observation is insufficient to conclude the spin freezing transition is also present in the irradiated films, instead it hints that the parameters of the spin Hamiltonian that govern the low temperature transition are similar in the three systems. In the \mbt\ material family there are multiple magnetic transitions in this temperature range. They include ferromagnetic transitions in the (MnBi$_2$Te$_4$)$_n$(Bi$_2$Te$_3$)$_m$ class \cite{sahoo2024}, but also the predicted ordering transition of a single isolated \mbt\ layer \cite{otrokov2019b}. We note therefore in the main text, that the low temperature transition may be related to the intralayer interactions in the system, which is consistent with the observed magnitude of the magnetization.}

\newpage
\section{Additional Transport data and interpretation}\label{si:trans}

\begin{figure}[ht]
    \centering
    \includegraphics[width=0.8\linewidth]{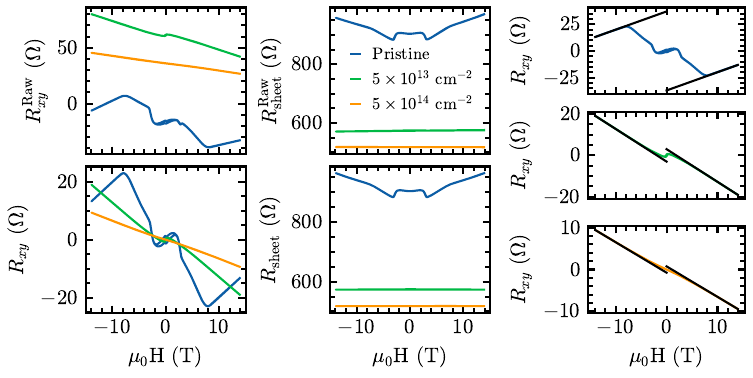}
    \caption{\add{Left and middle column gives the raw and (anti)symmetrized transverse $R_{xy}$ and sheet $R_{xx}$ resistance as a function of field. Right hand side column gives antisymmetrized $R_{xy}$ with linear OHE fitted at high field.}}
    \label{si:fig:sym}
\end{figure}

\subsection{Symmetrization and OHE}
\add{
The measured (raw) $R_{xx}$ and $R_{xy}$ curves are symmetrized by standard procedure. Two branches are measured (i and ii) with increasing and decreasing magnetic field. Transverse resistance is antisymmetrized as $R_{xy} = (R_{xy}^{\mathrm{i}}(H) - R_{xy}^{\mathrm{ii}}(-H))/2$ and longitudinal resistance is symmetrized as $R_{xx} = (R_{xx}^{\mathrm{i}}(H) + R_{xx}^{\mathrm{ii}}(-H))/2$.
The ordinary Hall effect is fitted to the symmetrized $R_{xy}$ by fitting a linear slope to the high field ($|\mu_0 H| > 10$~T) regions. Raw, (anti)symmetrized data, and OHE fits are given in Fig. \ref{si:fig:sym}.
}

\add{
We see no experimental evidence of multi-channel transport. At high fields the transverse resistance is linear. Values for the mobility and carrier density are calculated using the single channel Drude model as $n_{\mathrm{2D}} = 1/eR_0$ with $R_0$ the extracted ordinary Hall coefficient and $\mu = R_0/R_{\mathrm{sheet}}$ with $R_{\mathrm{sheet}} = R_{xx}(0~\mathrm{T}) \times (W/L)$. The mobility for the as-prepared, low fluence irradiated, and high fluence irradiated films are 17.3(3), 27.6(3), and 14.0(2) cm$^2$/Vs, respectively. Three dimensional carrier density is calculated as $n_{\mathrm{3D}} = n_{\mathrm{2D}}/t$ with $t = 75$~nm, accounting for the full stack: buffer, film, and capping. The carrier density for the as-prepared, low fluence irradiated, and high fluence irradiated films are $4.87\times10^{19}$, $5.26\times10^{19}$, and $1.14\times10^{20}$~cm$^{-3}$, respectively. 
Both carrier density and mobility values for the as-prepared film are in line with literature values for metallic MnBi$_2$Te$_4$ flakes \cite{cui2019}. Mobility values across the irradiation series remain in the same order of magnitude, with a slight increase from the as-prepared to low-fluence irradiated film.}

\subsection{Parallel Transport by Buffer Layer}
\add{
In the pristine and low fluence irradiated sample a well defined buffer layer is present in the stack. After high fluence irradiation, this (originally Bi$_2$Te$_3$) buffer evolves to the same disordered crystal structure as the MnBi$_2$Te$_4$ film as indicated by XRD. The buffer layer may therefore influence the Hall response of the pristine and low fluence irradiated film. We attribute the change in majority carrier type to a change in the electronic properties of the MnBi$_2$Te$_4$ film. A counter argument may be that the buffer dominates or strongly contaminates the Hall signal and the change of majority carriers is an effect of the buffer, not the MnBi$_2$Te$_4$ film. This requires, however, the MnBi$_2$Te$_4$ and buffer layer to carry opposite majority carriers, which introduces a depletion layer between the buffer and film. The Hall mobilities and carrier densities lie in the regime of metallic MnBi$_2$Te$_4$ and Bi$_2$Te$_3$ film with carrier densities typically reported in the order of $10^{19}$--$10^{20}$~cm$^{-3}$. A dielectric constant of $\varepsilon_r = 50$ has been reported for \bt\ \cite{richter1977} (see Table V therein), which we also use for \mbt\ (same material family). With the Bi$_2$Te$_3$ bandgap (165~meV \cite{chen2009}) as a lower bound for $\Delta V$ we estimate the width of such a depletion layer to be $W = \sqrt{(2\varepsilon_0\varepsilon_r \Delta V/q)(1/n_A + 1/n_d)} \sim$ {{$4$--$13$~nm}}. Here $\varepsilon_0$ is the vacuum permittivity, $\varepsilon_r$ the relative permittivity, $\Delta V$ the potential across the depletion layer, $q$ the elementary charge, and $n_A$ and $n_d$ the acceptor and donor (hole and electron) carrier densities. The width is the same order of magnitude as the buffer layer thickness of 10~nm. The requirement for the change in carrier type to be attributed to the buffer layer requires the existence of a depletion layer that nullifies the transport through the buffer layer. We conclude the observed change in Hall slope can not occur without a change in majority carrier type in the \mbt\ film.
}
\subsection{Uncertainty estimates}
\add{
The confidence in resistance, conductivity values, and extracted parameters is influenced by measurement conditions and data analysis. The instrumental uncertainty on both $L$ and $W$ is 0.1~mm. We propagate this uncertainty to the conductivity by error propagation on the geometric factor ($W/L$) to estimate the error on $\rho_{xx}$. The error on $R_{\mathrm{AHE}}$ is established through the fitting of the OHE at high field values, and establishes the error on $\rho_{xy}^{\mathrm{A}}$. The estimated standard deviation is then calculated through the tensor inversion and provides error estimates on $\sigma_{xx}$, $\sigma_{xy}$, and $\sigma_{xy}^{\mathrm{A}}$.
The irradiation-induced changes in $\sigma_{xx}$ and $\sigma_{xy}^{\mathrm{A}}$ are statistically significant with regards to the error estimates. Errors on parameter values extracted from transport propagate both the error from subtraction of the OHE and the geometric factor where relevant.
}
\subsection{Resistance versus temperature}
\begin{figure}[ht]
    \centering
    \includegraphics[width=0.75\linewidth]{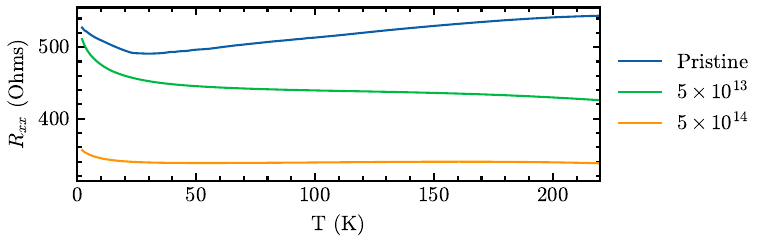}
    \caption{\add{The temperature dependence of the sheet resistance $R_{xx}$ measured over the irradiation series at $\mu_0H=0$~T. The $R_{xx}$--T curves were measured after storage of the samples in ambient conditions, resulting in absolute resistance values that differ from those in Fig. \ref{si:fig:sym}. The trend in resistance with fluence is unaffected.}}
    \label{si:fig:RT}
\end{figure}

\add{
Figure \ref{si:fig:RT} shows the sheet resistance $R_{xx}$ measured as a function of temperature. The as-prepared film's behavior is consistent with a bad metal, with changes in $R_{xx}$ up to 10 \% over the measured temperature range. No clear anomaly is observed at $T_{\text{N}}$, the magnetic order is instead evidenced by the anomalous Hall effect (Fig. \ref{fig:trans}). After irradiation, the $R_{xx}$-T curves are indicative of a film on the edge between metallic and insulating behavior, with thermally activated scattering playing a minor role in electrical transport. In all films an increase in resistance is observed at low temperatures, which may be attributed to static disorder from antisite defects (also present in the pristine film).
}

\end{document}